\documentclass{sn-jnl}
\usepackage[caption=false]{subfig}
\usepackage{graphicx}
\usepackage{color}

\title[Runtime Support for Performance Portability on Distributed Platforms]{Runtime Support for Performance Portability on Heterogeneous Distributed Platforms}
\author*[1]{\fnm{Polykarpos} \sur{Thomadakis}}\email{pthom001@odu.edu}
\author[1]{\fnm{Nikos} \sur{Chrisochoides}}\email{nikos@cs.odu.edu}

\affil[1]{\orgdiv{Department of Computer Science}, \orgname{Old Dominion University}, \orgaddress{\city{Norfolk}, \postcode{23529}, \state{Virginia}, \country{USA}}}

\begin{document}

\abstract{
Hardware heterogeneity is here to stay for high-performance computing. Large-scale systems are currently equipped with multiple GPU accelerators per compute node and are expected to incorporate more specialized hardware. This shift in the computing ecosystem offers many opportunities for performance improvement; however, it also increases the complexity of programming for such architectures. This work introduces a runtime framework that enables effortless programming for heterogeneous systems while efficiently utilizing hardware resources. The framework is integrated within a distributed and scalable runtime system to facilitate performance portability across heterogeneous nodes.
Along with the design, this paper describes the implementation and optimizations performed, achieving up to 300\% improvement on a single device and linear scalability on a node equipped with four GPUs.  The framework in a distributed memory environment offers portable abstractions that enable efficient inter-node communication among devices with varying capabilities. It delivers superior performance compared to MPI+CUDA by up to 20\% for large messages while keeping the overheads for small messages within 10\%. Furthermore, the results of our performance evaluation in a distributed Jacobi proxy application demonstrate that our software imposes minimal overhead and achieves a performance improvement of up to 40\%. This is accomplished by the optimizations at the library level as well as by creating opportunities to leverage application-specific optimizations like over-decomposition.
}

\maketitle

\section{Introduction}
\label{sec:intro}

The recent slowdown in Moore's Law is leading to large-scale disruptions in the computing ecosystem. Users and vendors are transitioning from utilizing computing nodes of relatively homogeneous CPU architectures to systems led by multiple GPU devices per node. This trend is expected to continue in the foreseeable future, incorporating many more types of heterogeneous devices, including FPGAs, System-on-Chips (SoCs), and specialized hardware for artificial intelligence\cite{osti_1822199}. The new computing ecosystem sets the basis to significantly improve performance, energy efficiency, reliability, and security; thus, high-performance computing (HPC) systems are adapted and optimized for traditional and modern workloads.     

Exploiting extreme heterogeneity requires new techniques and abstractions that handle the increasing complexity in productivity, portability, and performance. The new methods should allow users to express their applications' workflow uniformly, hiding the peculiarities of the underlying architecture while handling concerns arising from performance portability. One such concern is managing data on various devices. In most cases, data need to be transferred among devices to execute kernels optimized explicitly for an accelerator; thus, a framework needs to allocate the respective memory, find the devices involved in such a transaction, initiate the transfer, and monitor its progress. The coherence of the same data in different devices is also an issue. One needs to guarantee that the application will always use the most recent version of data, no matter which device. Moreover, since these operations are substantial overheads, they should happen asynchronously and overlap with valuable work, further increasing complexity.

Another concern is the orchestration of task (computation) execution. Tasks should only start asynchronous executions after the respective data have been moved to the target device and only when they do not conflict with other tasks, requiring lightweight synchronization. Task scheduling and load balancing should also be a significant concern to keep the available devices saturated and fully utilize them. The schedulers should be aware of each device's load and the data locality of each task to designate where each computation should occur efficiently. Finally, a framework that handles all these concerns should provide a friendly, high-level interface for applications but also expose low-level access that allows experts to optimize their applications for their specific needs. Moreover, having the option of lower-level access is crucial for distributed memory frameworks to use them on multiple nodes efficiently.

Utilizing and orchestrating data movement and task execution on multiple heterogeneous nodes increases the number of issues that need to be tackled. Thus, a complete runtime framework should also facilitate the seamless use of distributed heterogeneous nodes using the same approach of abstractions for data and workload independent of the underlying hardware and have tight integration with the performance portability layer used in a single node. Current trends in HPC follow the programming model of MPI+CUDA, which leads to complicated code, suboptimal performance, or both. Users that follow this approach need to explicitly transfer data between the host and the device before sending/receiving to/from a remote node. Moreover, they will need to use asynchronous operations for both memory transfer operations (host-GPU and network) and overlap them to avoid wasting cycles. This leads to the concern of correctness and synchronization involving the two types of data transfers and asynchronous kernel invocations. And even if one manages to handle all those correctly, they would have an application that only operates efficiently (or at all) for the specific hardware it was developed. Thus, a distributed framework that natively incorporates and abstracts heterogeneous nodes is the only way to create applications that scale independently of the hardware in which they were implemented.

In~\cite{Thomadakis22M}, we presented the Parallel Runtime Environment for Multicomputer Applications (PREMA), a scalable runtime system for distributed homogeneous platforms. It uses high-level abstractions to simplify distributed programming for dynamic and irregular applications. In this work, we extend PREMA to support seamless, efficient, and performance-portable development of distributed applications on heterogeneous nodes. First, we introduce a heterogeneous tasking framework to optimize the parallel execution of heterogeneous tasks on a single node. The tasking framework provides a programming model that automatically leverages heterogeneous devices. In contrast to other systems, our framework does not require the application to choose a device where a task should run; instead, the application only picks a device type, and the framework is responsible for scheduling the task to the optimal computing device. Next, we integrate PREMA with the heterogeneous tasking framework and enable it to manage and utilize heterogeneous nodes uniformly. 
Along with the design and implementation of the final product, we present optimizations that contribute to achieving high performance. The evaluation results with microbenchmarks and a proxy application show that our system incurs low overhead with scalable performance.

\subsection{Parallel Runtime Environment for Multicore Applications}
\label{sec:prema_bg}

\begin{figure}
    \centering
    \includegraphics[width=.8\linewidth]{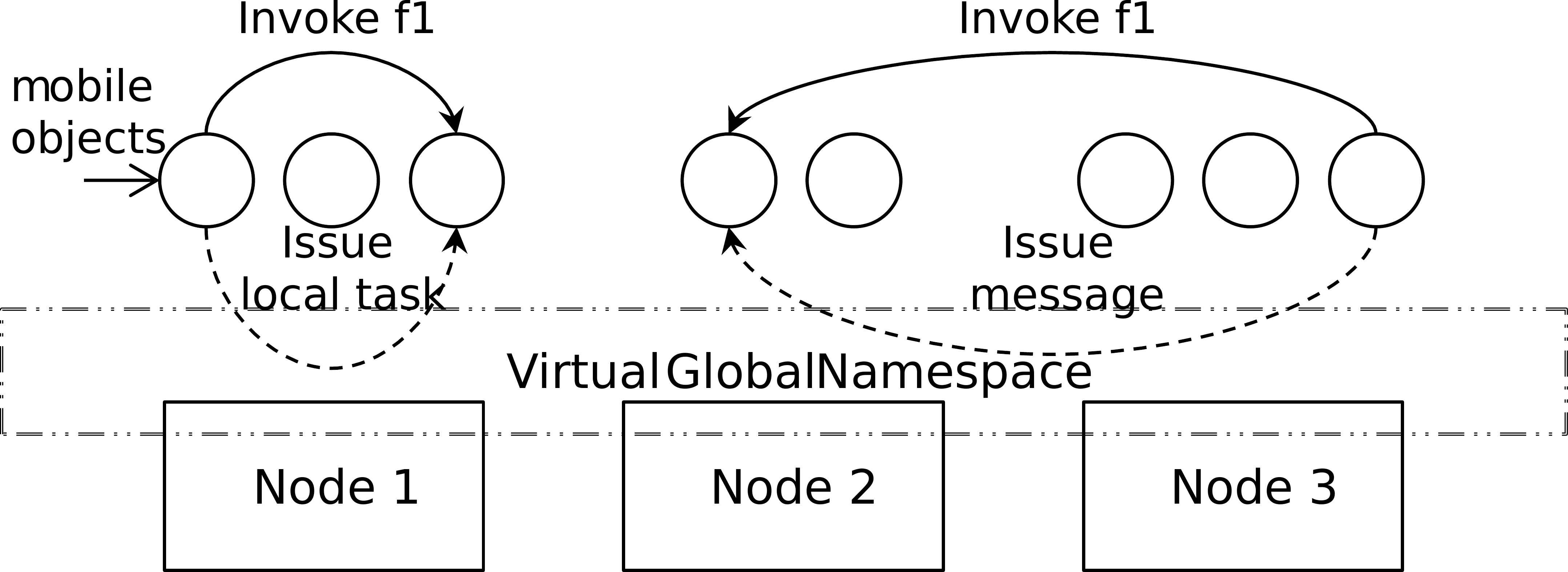}
    \caption{PREMA's mobile object driven (MOD) model. Applications are expressed as method invocations between local or remote mobile objects. A virtual global namespace provides a uniform high-level interface to issue local or remote work. The runtime system is responsible for running a task locally or sending an active message  depending on the location of the target mobile object. Figure adapted from~\cite{Thomadakis23T}.}
    \label{fig:mod}
\end{figure}

PREMA is a system designed to provide runtime support for large-scale computing clusters. It utilizes a 2-level parallelism approach that employs Message Passing Interface (MPI) for inter-node communication and Pthreads or Argobots\cite{SeoArgobots} for intra-node coordination. 
The system uses the construct of mobile objects, which are globally addressable, location-independent containers that hold application data. Mobile objects enable the mobile object-driven (MOD) programming model, which facilitates interactions between local or remote mobile objects through remote method invocations (called handlers). Handlers can be invoked on mobile objects uniformly, regardless of whether their data are local or remote. This approach abstracts the complexity of work scheduling, load balancing, and communication overhead, allowing applications to utilize available computing power without explicit concurrency handling. Figure~\ref{fig:mod} shows an example of the MOD model. 

PREMA extracts shared-memory parallelism by running non-conflicting handlers concurrently while implicitly migrating mobile objects among computing nodes to provide distributed-memory load balancing. Thus, the hardware memory spaces and processing elements are virtualized, allowing inter-handler parallelism (multiple handlers running in parallel) and sharing mobile object workload for threads in the same node.
Additionally, it offers a module for easy experimentation and development of new 2-level load balancing/scheduling policies.

\subsection{Contributions}
\label{sec:cont}
This paper presents an effort to address the challenges of efficiently utilizing distributed heterogeneous computing nodes in a portable and performant way. It introduces a heterogeneous tasking framework as an extensible layer of portability. In addition, it presents the integration of the framework within a distributed memory system to produce a design that natively handles distributed nodes equipped with heterogeneous devices. 
The major contributions of this paper are as follows.
\begin{itemize}
    \item A new design and implementation of a heterogeneous tasking framework for the development of performance portable applications that can scale from single-core to multi-core, multi-device (\textit{CPUs, GPUs}) platforms efficiently, \textit{without any code refactoring}. 
    \item A novel integration of a distributed runtime with the heterogeneous tasking framework to provide an end-to-end solution that scales over distributed heterogeneous computing nodes while exposing a high-level and abstract programming model.
    \item A series of memory, scheduling, and threading performance optimizations that achieve significant improvements, up to 300\% on a single GPU and linear scalability on a multi-GPU platform, that are directly applicable to similar systems and applications.
    \item Demonstration of up to 40\% speedup on an end-to-end distributed, heterogeneous proxy application (e.g., Jacobi solver)  by utilizing the new runtime framework in combination with widely studied optimizations like over-decomposition\cite{chrisochoides1996multithreaded}.
    
\end{itemize}

\section{Related Work}
\label{sec:related_work}
Several systems have been adapted to efficiently utilize GPUs in their workflow, while new ones have emerged trying to create new standards for their use. Systems like Charm++~\cite{Charm++}, HPX~\cite{HPX}, and X10~\cite{X10} have introduced new interfaces to provide support for GPUs. However, these systems let the users explicitly handle issues like requesting memory transfers, managing device platforms, task allocations, and work queues to optimize performance. In contrast, the proposed work provides a uniform abstraction for heterogeneous tasks and data and implicitly handles scheduling, load balancing, and latency overlapping independently of the target device backend. StarPU~\cite{StarPU}, OmpSs~\cite{ompss}, and ParSec~\cite{ParSec} offer different high-level approaches for efficiently utilizing distributed heterogeneous systems. However, their programming model is better suited for applications whose workflow follows a regular pattern that can be inferred mostly statically.
On the other hand, PREMA adopts a dynamic, message-driven programming model that is more suitable for irregular applications. SYCL and DPC++/oneAPI~\cite{DPC++}, as well as the newest version of OpenMP, are recent attempts to provide performance portable interfaces in modern C++ that can target heterogeneous devices. However, users still need to handle load balancing, scheduling, and work queues for multi-device systems and need to combine them with another runtime solution that targets distributed nodes.

\section{Design and Implementation}
\label{design}

\begin{figure}
    \centering
    \includegraphics{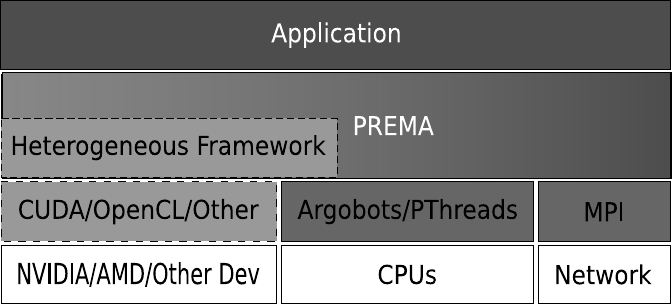}
    \caption{A high-level representation of the heterogeneity-aware PREMA. The hardware devices/interfaces stand on the lower level and are utilized by integrating PREMA with MPI, PThreads, and Argobots (CPU-only; see~\cite{Thomadakis22M},~\cite{Thomadakis23T}), and the heterogeneous tasking framework (in the current work). On top of that stands the application, which leverages these capabilities through a simple but powerful interface.}
    \label{fig:h_prema}
\end{figure}

Following the principle of separation of concerns, a new abstract compatibility layer is introduced, allowing PREMA to access different heterogeneous devices uniformly. This layer is implemented as a stand-alone heterogeneous tasking framework that handles all concerns arising from the co-existence of multiple types of devices. PREMA integrates this framework as the preferred way to interact with heterogeneous devices. Thus, it can be easily extended to utilize more device types without needing to modify its implementation, apart from this low-level compatibility layer. Moreover, PREMA exposes some of these capabilities wrapped in a high-level interface, allowing users to utilize such devices in a controlled and safe way. Figure~\ref{fig:h_prema} shows a high-level representation of the software stack and how the different layers interact. In the following sections, first, we present the heterogeneous framework layer and its capabilities in detail; then, we focus on its integration with PREMA to provide a distributed, heterogeneity-aware runtime.

\subsection{Heterogeneous Tasking Framework}
\definecolor{dkgreen}{rgb}{0,0.6,0}
\definecolor{gray}{rgb}{0.5,0.5,0.5}
\definecolor{mauve}{rgb}{0.58,0,0.82}
\lstset{frame=tb,
  language=C++,
  aboveskip=2mm,
  belowskip=2mm,
  showstringspaces=false,
  columns=flexible,
  basicstyle={\footnotesize\ttfamily},
  numbers=none,
  numberstyle=\footnotesize\color{gray},
  keywordstyle=\color{blue},
  commentstyle=\color{dkgreen},
  stringstyle=\color{mauve},
  breaklines=true,
  breakatwhitespace=true,
  tabsize=3
}

\begin{figure}[!h]
    \centering
    \begin{lstlisting}[numbers=left]

// Device-independent kernel implementation
kernel(dgemm, (device_global double* A, device_global double* B, device_global double* C, long N),
{
    parallel_region(
        int ROW = kernel_group_id_y * kernel_local_size_y + kernel_local_id_y;
        int COL = kernel_group_id_x * kernel_local_size_x + kernel_local_id_x;
    
        double local_sum = 0;
        
        if (ROW < N && COL < N) 
        {
            for (int i = 0; i < N; i++) 
            {
                local_sum += A[ROW * N + i] * B[i * N + COL];
            }    
            
            C[ROW * N + COL] = local_sum;
        }
    )
})
    
int main()
{
    const int N = 1024;
    
    // Allocate NxN matrices A, B, C 
    hetero_object<double> m_A(N,N);
    hetero_object<double> m_B(N,N);
    hetero_object<double> m_C(N,N);

    // Get write access to data and populate them
    double *A = m_A.request_data(false, true).get();
    double *B = m_A.request_data(false, true).get();
    ...

    // Release data access from host
    m_A.release();
    m_B.release();

    {
        hetero_task task;
        
        // Set the input/output matrices A, B, C
        task.arg(m_A).read();
        task.arg(m_B).read();
        task.arg(m_C).write().dim_x();

        // Set the thread dimensions
        task.set_threads({32, 32, 1}, {32, 32, 1});

        // Set the target device type
        task.device(device::GPU);
        
        // Set the kernel to execute
        task.submit(dgemm);
    }
}
    \end{lstlisting}
    \caption{An example of a DGEMM application using the tasking framework.}
\label{fig:dgemm}
\end{figure}

    
    
        
        

The programming model of the heterogeneous tasking framework builds upon two simple abstractions: the heterogeneous objects (\emph{hetero\_objects}) and heterogeneous tasks (\emph{hetero\_tasks}). A \emph{hetero\_object} uniformly represents a user-defined data object residing on one or more computing devices of a heterogeneous compute node (e.g., CPUs, GPUs, FPGAs). Applications treat such objects as opaque containers for data without being aware of their physical location. A \emph{hetero\_task} encapsulates a non-preemptive computing kernel that runs to completion and implements a medium-grained parallel computation. Like hetero\_objects, hetero\_tasks are defined and handled by the application uniformly, independent of the device they will execute on.
Fig.~\ref{fig:dgemm} shows an example of a DGEMM implementation using the tasking framework. The presented example shows a DGEMM execution request for the GPU; however, by only changing the device target in line 53, one can target a different device without touching the rest of the code. The kernel() macro on the top of the listing will expand to CUDA, OpenCL, and/or other defined backends to fit the vendor-provided implementation.

\subsubsection{Heterogeneous Objects} 
Handling copies of the same data on different heterogeneous devices can lead to error-prone and difficult-to-maintain application code. In general, applications need to manage data transfers among them, use the correct pointer for the respective device, and keep track of their coherence. A  hetero\_object is an abstraction that automatically handles such concerns, maintaining the different copies of the same data in a single reference. The underlying system controls hetero\_objects to guarantee that the most recent version of the data will be available at the target device when needed. For example, accessing an object currently resident on the CPU from a  GPU would automatically trigger the transfer of the underlying data from the host to the respective device. In the same manner, accessing the same object from a different device would initiate a transfer from the GPU to that device, potentially after first staging the data at the host. Finally, the runtime system guarantees data coherence among computing devices, keeping track of up-to-date or stale copies and handling them appropriately.

The memory captured by a hetero\_object should mainly be accessed and modified through hetero\_tasks for optimal performance. However, the application can also explicitly request access to the underlying data on the host after specifying the type of access requested to maintain coherence. This method will trigger (if needed) an asynchronous transfer from the device with the most recent version of the data and immediately return a future. The future allows for querying the transfer status and providing access to the raw data once the transfer has been completed. In this state, the data of the hetero\_object are guaranteed to remain valid on the host side, preventing tasks that would alter them from executing until the user explicitly releases their control back to the runtime system. Since the application has no direct access to the memory allocated to different devices, our framework monitors the memory usage of each device. When a device's memory is close to being depleted, the runtime system will automatically start offloading some of the user's data to the host or other devices. We currently use a Least Recently Used (LRU) policy to determine which hetero\_object should be offloaded to free a device's memory. An application can explicitly request to remove a hetero\_object from all devices to help the runtime clean up some memory; otherwise, a hetero\_object will be freed when going out of scope. In both cases, the hetero\_object will only be removed once no tasks and other operations are referencing it. 

\subsubsection{Heterogeneous Tasks} 
Heterogeneous tasks (hetero\_tasks) are opaque structures that consolidate the parameters characterizing a computational task. Through a hetero\_task, applications define the kernel to execute, input/output data arguments, processing elements requested (e.g., threads in a CPU), task dependencies, and target device type. Moreover, applications can request the allocation of a temporary shared memory region available only for the duration of the kernel, which maps to the concept of local/shared memory found in other GPU programming APIs (CUDA, OpenCL).

Heterogeneous tasks are independent of the underlying target hardware, allowing a uniform expression of the application workflow whether they target CPUs, GPUs, or other device types. The computational kernel they represent is defined in a dialect similar to an OpenCL kernel that is translated appropriately for each target device. Input/output data arguments of a task are defined as the hetero\_objects it needs to access, along with the access type required for each (read, write, read-write). This information is used to issue the appropriate data transfers, maintain coherence, and infer task dependencies. Submitting a task for execution does not immediately execute the respective kernel; instead, the runtime system enqueues the task execution request and immediately returns control to the user. The heterogeneous tasking framework provides methods to query the status of a kernel execution or wait for its completion. Moreover, task dependencies can be defined either explicitly by the user or implicitly by the runtime.

Applications can \textbf{explicitly define a task dependency graph} using the \emph{add\_dependency()} method of hetero\_tasks. Once their dependencies have been set, the application can submit the respective tasks for execution all at once. 
This approach allows the runtime system to improve performance while removing much of the burden of guaranteeing correctness from the application. 
To further reduce the effort required to guarantee correctness, the framework also supports \textbf{implicit task dependency detection} based on the arguments accessed by each hetero\_task. Assuming that the application submits tasks in the correct sequential order, conflicting tasks are guaranteed to execute in the proper order; independent tasks will automatically explore maximum parallelism and avoid race conditions.


\subsubsection{Execution Model}
Heterogeneous tasks are executed asynchronously by the tasking framework. A task submitted for execution is appended to a list of task execution requests. The control is then immediately returned to the application, which can continue to issue more tasks or execute other work. A separate component of the runtime (optionally running in a separate thread) examines task execution requests and eventually schedules them for execution after performing the necessary steps to guarantee correctness.

The first step towards executing a task is to infer its dependencies with other tasks based on their data arguments. The runtime maintains a list of the currently submitted or running tasks that target each hetero\_object;  new tasks that access these hetero\_objects with a conflicting access type have their dependencies set accordingly. Those with at least one incomplete dependence are pushed to another queue of blocked tasks; otherwise, they are appended directly to the scheduler's runnable work pool. Blocked tasks are periodically checked for resolved dependencies, and those with all their dependencies resolved are moved to the scheduler's runnable tasks pool.

Once all blocked tasks have been examined, the scheduler is ready to schedule the runnable tasks. At this stage, the scheduler decides the order in which the tasks should execute and the device where they should run based on the user's device type preference (i.e., the scheduler chooses the specific device ID while the user only gives a selection for a device type). The runtime will then reserve device resources and issue the data transfer requests of the input/output hetero\_objects to be accessed on the chosen device. Moreover, it will automatically try to overlap the different operations, if possible, by utilizing the features provided by the target device's API (e.g., CUDA streams or OpenCL command queues). When all outstanding data transfers of a task have been completed, the computational kernel will be submitted to the target's work pool. Submitted tasks are periodically checked for completion by the runtime to update the status of pending dependent tasks.

\subsubsection{Scheduler}
With the introduction of more heterogeneous computing devices and workloads, it is expected that scheduling and load balancing will only become more complicated. To provide flexibility for different use cases, the actual implementation of the scheduler is designed to be modular and separate from the rest of the heterogeneous tasking framework. We provide the scheduler as an abstract class that only requires two operations to be implemented. The \emph{push()} operation adds a new runnable task into the scheduler's work pool while the \emph{pop()} operation returns the next task to be executed as well as the device it should run on. The abstract scheduler class allows the development of as simple or complex custom data structures and policies as the user might need. 

\subsubsection{Implementation}
\begin{figure}
    \centering
    \includegraphics[width=.7\linewidth]{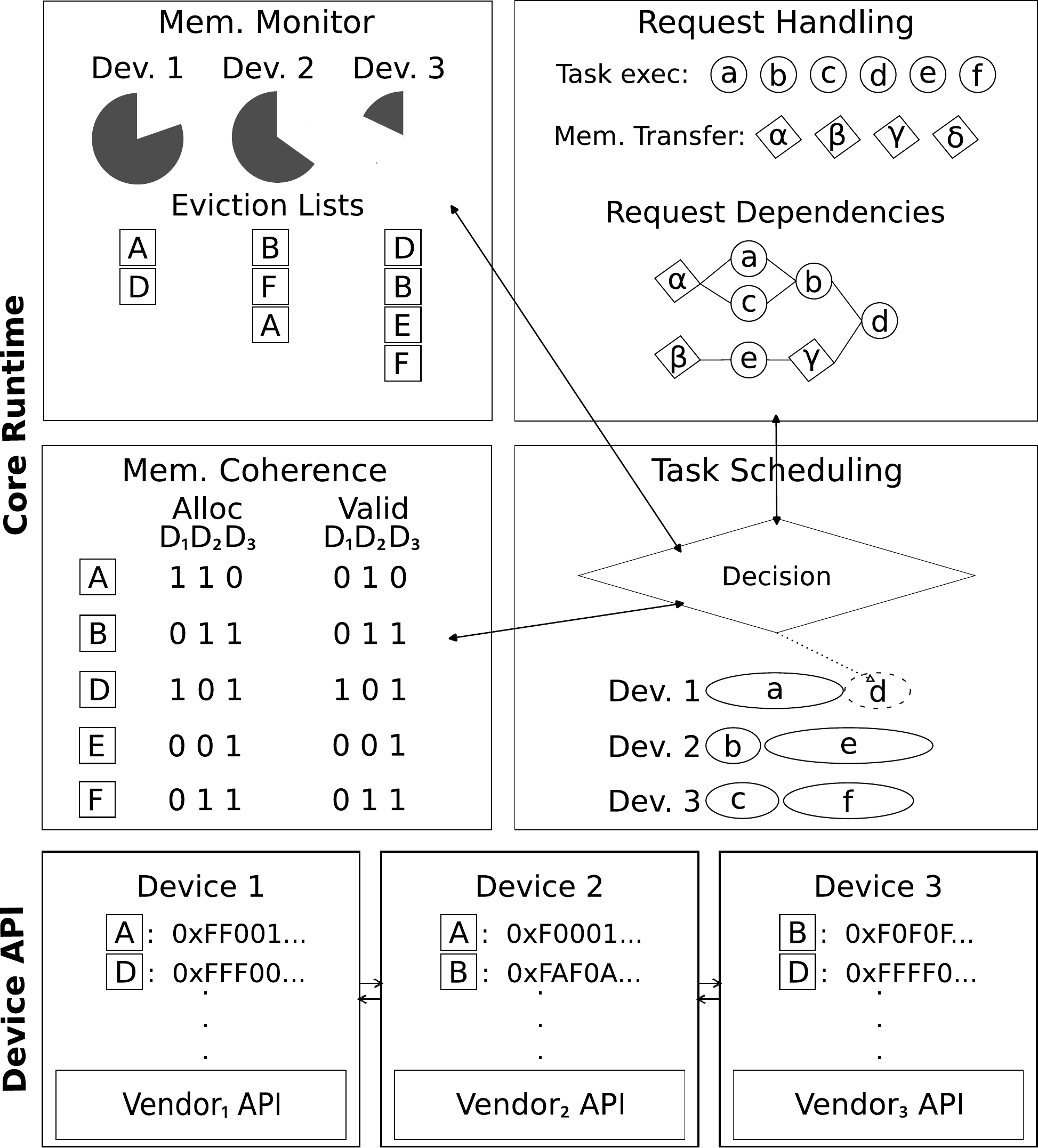}
    \caption{A high-level representation of the heterogeneous tasking framework software stack and operations. The operations performed by the Core Runtime include (a) Memory monitoring to keep track of the available device memory and deallocate unused objects when running low on resources, (b) Memory transfer and task execution request handling that dispatches such requests when it is safe, (c) Memory coherence among different copies of the same hetero\_object in multiple devices, (d) Task scheduling to optimize for a reduction in memory transfers and optimize overall execution time. The Device API exposes an abstract ``device class'' that encapsulates the implementation of different vendor interfaces uniformly. The device API maps high-level abstractions like hetero\_objects and hetero\_tasks to actual device-specific constructs.}
    \label{fig:tf_overview}
\end{figure}
The heterogeneous tasking framework is implemented in the C++ programming language leveraging its performance and object-oriented design. It is developed in three software layers to allow easy integration with new device types, programming APIs, and scheduling policies (see Fig.~\ref{fig:tf_overview}).


The \textbf{Device API} is the bottom layer, encapsulating the different operations provided by a heterogeneous device vendor. It consists of abstract C++ classes that expose virtual methods for operations required to (a)synchronously issue tasks and manage data in such devices, query their hardware specifications, and methods to retrieve the status of an asynchronous operation. Currently, we provide native support with CUDA and OpenCL for GPUs. The Device API provides the low-level, vendor-specific implementations of all abstractions of the tasking framework, like mapping hetero\_objects to memory locations, hetero\_tasks, and task execution requests to actual kernel invocations, memory transfer requests, etc. 


The next layer is the \textbf{Core Runtime} layer, which provides the underlying implementation of the hetero\_objects and hetero\_tasks, monitors the coherence of the different copies of the data, and detects and enforces task dependencies. It utilizes the Device API to coordinate data transfers, guide the correct execution of tasks and signal the completion of different operations.
This layer acts as the ``glue'' between the application preferences, the scheduler and load balancing policies, and the Device API.

At the top stands the \textbf{Application Layer}, which consists of a thin API that exposes the capabilities of the tasking framework in a high-level interface. 
In the current implementation, kernels are defined in a dialect similar to OpenCL through the use of macros which are expanded to implement a kernel version for each available target. 

\subsection{Heterogeneity within PREMA}
\label{sec:prema_integration}

Integrating heterogeneity in PREMA is a crucial requirement to handle the load of exascale-era machines. Applications should be able to use and transfer device memory in the context of remote handler executions without much hassle. A step towards this direction is to allow PREMA to send and receive buffers located in a GPU device either explicitly (currently CUDA only) or through the abstractions of the heterogeneous tasking framework we have introduced. The explicit approach allows users to utilize GPUs without confining them to use our heterogeneous tasking framework, facilitating interoperability with legacy CUDA codes. It also provides a barebone approach to integrate heterogeneity on top of the distributed system, which can act as the base case for our performance evaluation since it introduces the least possible overhead.

\subsubsection{Explicitly Handling Devices}
In the explicit approach, the application can directly call the different GPU operations of the CUDA API to allocate/free memory, initiate transfers and execute kernels. PREMA provides a function to invoke remote handlers that include a GPU buffer as an argument; the function requests the ID of the remote process, the buffer to transfer, its size, the IDs of the source and target devices, and the handler (host function) to be invoked at the receiver. PREMA will transfer the buffer between the remote GPUs and invoke the handler when it has been completed. The handler can then invoke any GPU-related operation that targets this buffer safely. However, the application needs to guarantee that the handler does not return before the completion of the kernel since any buffers transferred through a handler will be freed at its return, including the GPU buffer. Waiting for all the device operations to complete (e.g., through \emph{cudaDeviceSynchronize()}) is enough to guarantee correctness; however, this approach will harm PREMA's time-slicing abilities, preventing it from switching to other tasks while GPU operations are in progress. Thus, the user should follow a more complicated approach, querying the status of the operations without blocking (e.g., through \emph{cudaEvents}) and periodically yielding control of the thread for PREMA to run background jobs.

\subsubsection{Utilizing the Heterogeneous Tasking Framework}
\definecolor{dkgreen}{rgb}{0,0.6,0}
\definecolor{gray}{rgb}{0.5,0.5,0.5}
\definecolor{mauve}{rgb}{0.58,0,0.82}
\lstset{frame=tb,
  language=C++,
  aboveskip=2mm,
  belowskip=2mm,
  showstringspaces=false,
  columns=flexible,
  basicstyle={\footnotesize\ttfamily},
  numbers=none,
  numberstyle=\footnotesize\color{gray},
  keywordstyle=\color{blue},
  commentstyle=\color{dkgreen},
  stringstyle=\color{mauve},
  breaklines=true,
  breakatwhitespace=true,
  tabsize=3
}

\begin{figure}[!h]
    \centering
    \begin{lstlisting}[numbers=left]

// PREMA mobile object handler 
DEFINE_MP_HANDLER(execute_second_dgemm_handler)
{
    hetero_object<double> m_B = get_hetero_object<double>();
    hetero_task task;
    
    // Set the input/output matrices A, B, C
    task.arg(this->m_A).read();
    task.arg(m_B).read();
    task.arg(this->m_C).write().dim_x();
    task.set_threads({32, 32, 1}, {32, 32, 1});
    task.device(device::GPU);
    task.submit(dgemm);
}

// PREMA mobile object handler 
DEFINE_MP_HANDLER(execute_first_dgemm_handler)
{
    hetero_task task;
    hetero_object m_C;
    task.arg(this->m_A).read();
    task.arg(this->m_B).read();
    task.arg(m_C).write().dim_x();
    task.set_threads({32, 32, 1}, {32, 32, 1});
    task.device(device::GPU);
    task.submit(dgemm);

    prema::mp_send(this->other_mp, execute_second_dgemm_handler, m_C);
}
    
    
int main()
{
    prema::init();
    const int N = 1024;
    
    // Allocate NxN matrices A, B, C 
    hetero_object<double> m_A(N,N);
    hetero_object<double> m_B(N,N);
    hetero_object<double> m_C(N,N);

    // Populate A, B 
    ...

    mobile_object_data my_data(m_A, m_B, m_C);
    
    // my_mp is a reference to the remote mobile object
    prema::mobile_ptr my_mp(my_data);
    
    // other_mp is a reference to the remote mobile object
    prema::mobile_ptr other_mp = /*get remote mobile_ptr*/;
    my_data.other_mp = other_mp;    
    
    prema::mp_send(my_mp, execute_first_dgemm_handler);

    prema::shutdown();
}
    \end{lstlisting}
    \caption{An example of a series of two DGEMM invocations where the results of the first are needed on a remote node to invoke the second, using the heterogeneity-aware PREMA.}
\label{fig:dist_dgemm}
\end{figure}

    
    
        
        

To facilitate a higher-level interaction of PREMA with heterogeneous devices, we introduced a set of extensions allowing direct utilization of the abstractions provided by the heterogeneous tasking framework. 
Compared to the explicit remote handler invocation API, the user only needs to provide the handler to be executed, the target process ID, and the hetero\_object passed as an argument (transferred). 
Since the hetero\_objects handle the location of the underlying data, the user does not need to specify their location. The framework automatically decides the device to store the received buffer on the target process. Once a hetero\_object of a remote method invocation has been transferred, the designated handler is invoked on the target. The application can invoke tasks that utilize it on any available device type. Moreover, the application is guaranteed that any hetero\_object that is the target of any hetero\_task execution or messaging operation will live long enough for all such operations to complete, even if the handler returns earlier. In addition, the tasking framework will make sure that no other task can start executing on a hetero\_object that is in the process of network transfer. A code example is shown in Fig.~\ref{fig:dist_dgemm} where a series of two distributed DGEMM invocations is implemented in heterogeneous PREMA. Two mobile objects create matrices A, B, and C, then create mobile pointers out of their data and exchange them with each other. Next, each mobile object invokes the first DGEMM on itself and sends the result to the remote object invoking the second DGEMM. Note that there is no need to explicitly handle the data buffers for the network transfer (lines 29, 55). Also, the application does not need to explicitly wait for the completion of the DGEMM task (line 27) before sending the results to the remote mobile object (line 29). Finally, even though the result of the first DGEMM is stored in a local variable (lines 20, 24) and the handler can return before the asynchronous send has completed, the data will be transferred correctly. PREMA and the tasking framework will make sure that data are consistent and updated in the correct order.  


Another desirable requirement provided through the hetero\_objects on top of PREMA is the ability to ``put'' and ``get'' data between potentially distributed devices. A new extension allows users to create global pointers for hetero\_objects, i.e., unique identifiers referenceable from all processes in the distributed system. 
When an application needs to store/retrieve data to/from a remote hetero\_object, it just needs to provide its global pointer and the location (hetero\_object or pointer) of the data to be read/written, along with a callback that is triggered on the target, signaling the completion of the operation.   

\subsubsection{Implementation}
\label{sec:hetero_prema_impl}

\begin{figure}
    \centering
    \includegraphics[width=\linewidth]{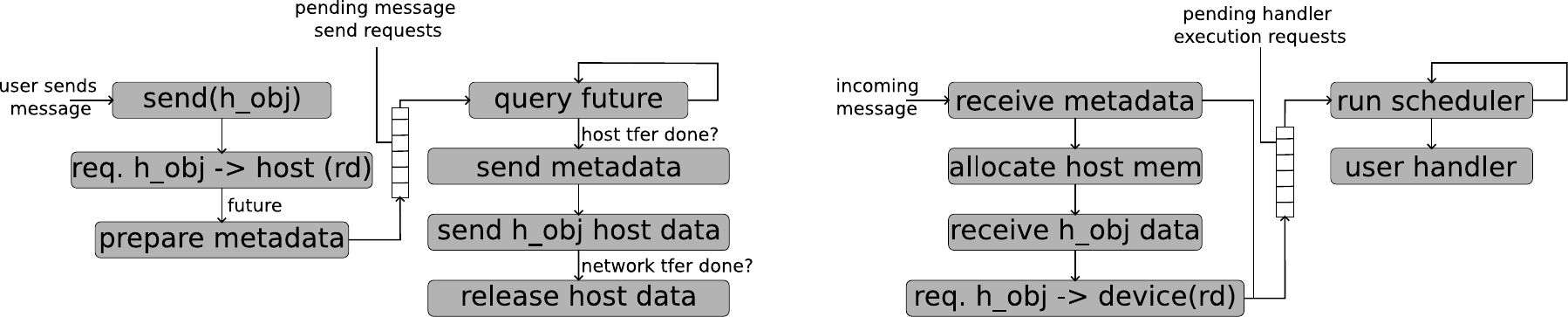}
    \caption{Heterogeneous data send (left) and receive (right) in PREMA with no GPU-aware interconnect.}
    \label{fig:prema_hetero_send_recv}
\end{figure}

Depending on the capabilities of the underlying communication library and the target device hardware, the actual implementation of the memory transfers differs to leverage heterogeneity-aware communication substrates.
When the application utilizes hetero\_objects, and the communication substrate is not heterogeneity-aware, the implementation of the memory transfers includes the following steps (also see Fig.~\ref{fig:prema_hetero_send_recv}):
\\

\textbf{Sender:}
\begin{enumerate}
    \item Asynchronous memory read from a device to the host returning a future.
    \item Push a message, including the future and handler metadata, to the outgoing message pending queue.
    \item Periodically query the outgoing message queue. 
    \item When the future of a message is complete, send two messages: the metadata and the hetero\_object.
    \item Once the message transmission is complete, release access to the host data.
\end{enumerate}

\textbf{Receiver:}
\begin{enumerate}
    \item Receive metadata.
    \item Use the information in metadata to prepare for the hetero\_object data.
    \item Receive the hetero\_object in the data prepared.
    \item Request allocation of the hetero\_object in a device.
    \item Execute the user handler.
\end{enumerate}

PREMA will automatically request an asynchronous read of the hetero\_object data from the device to the host. The tasking framework will guarantee that the device-to-host transfer will start once all previously submitted, conflicting tasks have finished and prevent any new ones from running before PREMA has finished its network transfer. Next, a header message encapsulating the handler's metadata (e.g., data pointer, target, etc.) is prepared and pushed to a queue of pending send message requests. The header will also incorporate a future returned by the previous step that allows for checking for the transfer's progress. The outgoing message requests queue is checked periodically by PREMA. When the future of a message signifies the completion of a device-to-host transfer, the message is ready to be sent through the network. PREMA will asynchronously send two messages, one for the metadata and one for the hetero\_object data, utilizing the MPI as the communication substrate. Finally, when the transmission of the two messages has been completed, PREMA will release its read request from the hetero\_object, notifying the tasking framework that the object can be safely modified or deleted by another task.

On the receiving side, once the first metadata message is detected and received, the information it carries is used to allocate the required memory to store the actual data of the hetero\_object. Next, the second message with the actual hetero\_object data is received in the newly allocated buffer. A request to allocate the received data in a device is sent to the tasking framework, and the metadata message along with this request is enqueued into a pending handler execution request queue. Finally, the scheduler will pick one of the pending handler execution requests and run the respective user handler. Note that in this case, PREMA does not wait for the host-to-device transfer to complete before starting the handler, allowing code independent of the hetero\_object itself to run, overlapping the transfer. Furthermore, since the data of the hetero\_object will be used through a hetero\_task, the tasking framework will ensure that any task execution will be delayed until the transfer has been completed. 

The ``put'' operation is almost identical, with the small difference that the receiver does not need to allocate new host memory. Since the hetero\_object already exists on the target, PREMA will request write access on the host side, and once access is granted, it will receive the data directly in the hetero\_object's host memory. Note that no transfer from device to host is performed since the data will be overwritten from the receive. The request will just guarantee that the network transfer will not conflict with any other task running on this hetero\_object.

\begin{figure}
    \centering
    \includegraphics[width=\linewidth]{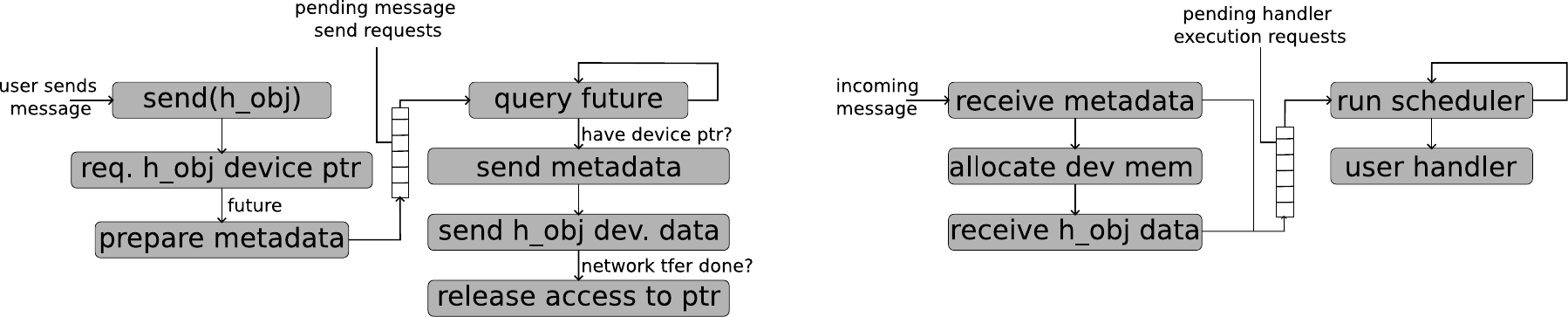}
    \caption{Heterogeneous data send (left) and receive (right) in PREMA with no GPU-aware interconnect.}
    \label{fig:prema_hetero_send_recv_dd}
\end{figure}

The host-staging step can be skipped if the communication library/hardware and compute devices support direct transfers between distributed devices (currently only tested for CUDA-OpenMPI). Thus, in an implementation for the case where the user uses hetero\_objects, the process is the following (also see Fig.~\ref{fig:prema_hetero_send_recv_dd}):
\\

\textbf{Sender:}
\begin{enumerate}
    \item Asynchronous request access to device memory returning a future.
    \item Push a message, including the future and handler metadata, to the outgoing message pending queue.
    \item Periodically query the outgoing message queue. 
    \item When the future of a message is complete, send two messages: the metadata from the host and the hetero\_object directly from the device.
    \item Once the message transmission is complete, release access to the device data.
\end{enumerate}

\textbf{Receiver:}
\begin{enumerate}
    \item Receive metadata. 
    \item Use the information in metadata to prepare for the hetero\_object data in the device.
    \item Receive the hetero\_object in the data prepared.
    \item Request allocation of the hetero\_object in a device.
    \item Execute the user handler.
\end{enumerate}

PREMA will automatically asynchronously request read access to the hetero\_objects pointer in the current device. Again, the tasking framework will ensure that no conflicts with active tasks will be possible, but no transfer will occur. Same as the non-GPU-aware case, a metadata message is prepared that includes a future generated from the previous step's request and is appended to the message send requests. Once it is safe for PREMA to access the device memory, the message is ready to be sent through the network. The metadata message is sent first, followed by the data in the device. Under specific conditions, some MPI implementations can directly target CUDA device memory which is used in this case. When the message transmissions have been completed, PREMA will release its read access to the hetero\_object. This will allow other task/messaging operations to modify the hetero\_object safely.

On the receiving side, once the metadata message is received, the tasking framework is requested to create a dummy hetero\_object on the device that can accommodate the incoming data. Then the metadata message along with the dummy hetero\_object are enqueued to the pending handlers queue for execution. When the hetero\_object allocation is complete, PREMA will receive the actual data directly in the memory allocated in the device and notify the tasking framework that it no longer needs to keep access to the device pointer. Finally, the scheduler will pick one of the pending handler execution requests and run it. Again, the handler can start before the actual data have been received in the hetero\_object, leaving the tasking framework to guarantee that no conflicts will occur by delaying tasks as needed. 

Like in the previous case, the ``put'' operation is, with the only difference that the receiver can directly receive incoming data on the current location of the hetero\_object. Since the hetero\_object already exists on the target, PREMA will request write access on its current device location. Once access is granted, it will receive the data directly in the hetero\_object's device memory. Again, data will be overwritten from the receive; thus, the request will guarantee that the network transfer will not conflict with any other task running on this hetero\_object.

\section{Performance Evaluation}
\label{sec:perf_eval}

This section investigates optimizations that can improve the performance of different operations provided by the heterogeneity-aware PREMA and the tasking framework and presents the performance of the optimized implementation on a proxy application. We used a small 16-node cluster, with each node consisting of two Intel Xeon Gold 6130 20-core CPUs (2.1 GHz) and four NVIDIA Tesla V100 GPUs.

\subsection{Heterogeneous Tasking Framework}
\begin{figure}[t]
    \centering
    \includegraphics[width=.7\linewidth]{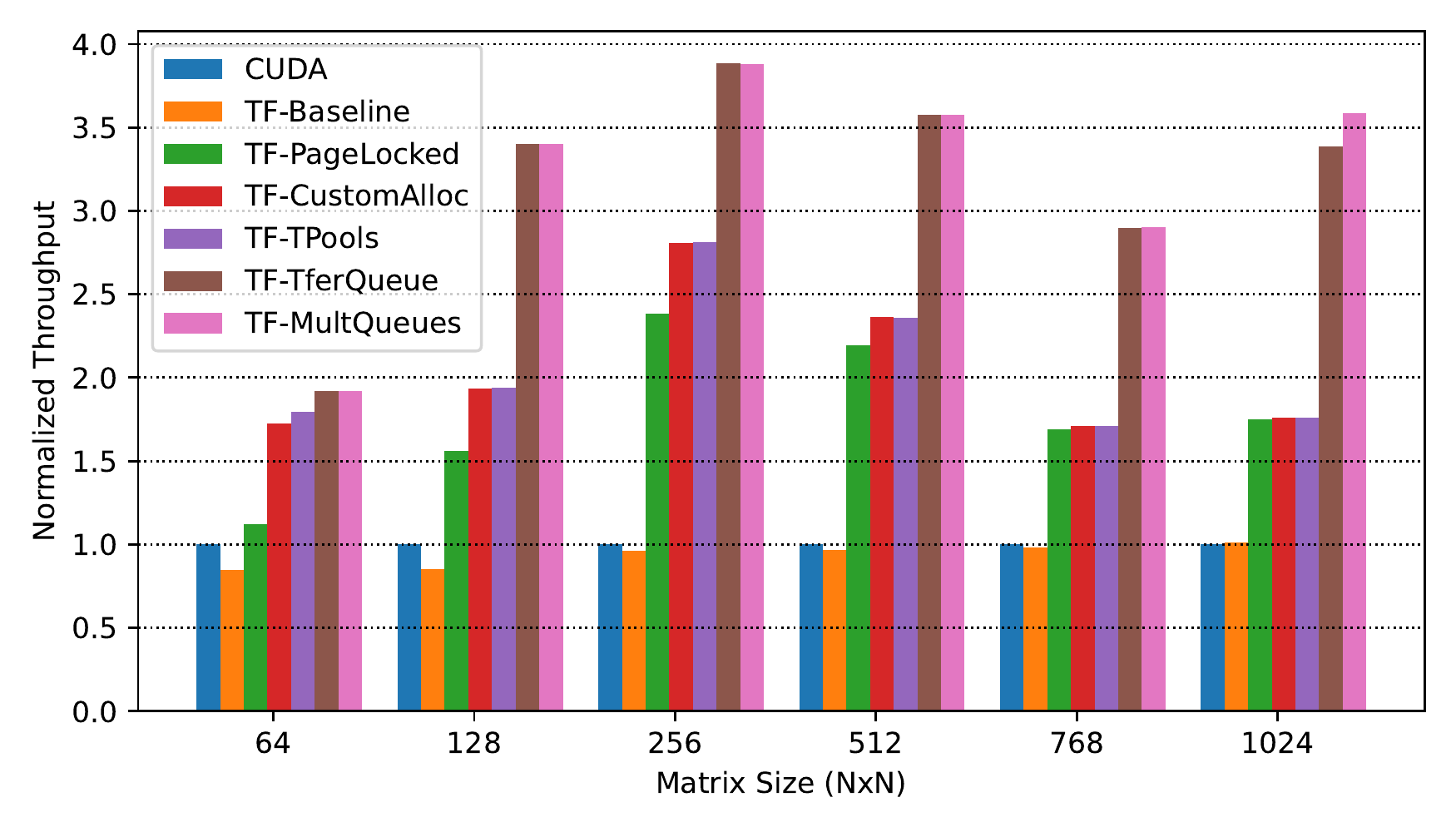}
    \caption{Effects of incrementally applying different optimizations on the heterogeneous tasking framework's performance for a matrix multiply benchmark on a single GPU. The performance improvements are evaluated on multiple matrix sizes and normalized by a CUDA implementation's performance. \textbf{CUDA}: The baseline CUDA implementation. \textbf{TF-Baseline}: The tasking framework (TF) without any optimizations. \textbf{TF-Pagelocked}: TF utilizing Page-locked host memory. \textbf{TF-CustomAlloc}: TF after applying the custom memory allocator optimization. \textbf{TF-TPools}: TF utilizing request pools. \textbf{TF-TferQueue}: TF after introducing a separate queue for memory transfers. \textbf{TF-MultQueue}: Final version of TF, utilizing multiple queues for kernel invocations.}
    \label{fig:dgemm_benchmarks}
\end{figure}

We evaluate different performance optimization techniques on this framework for NVIDIA GPUs using a simple, double precision matrix-matrix multiply benchmark and compare the throughput achieved in 100 iterations with a pure CUDA implementation. In each iteration, the three matrices are allocated in the device, input data are transferred, and the compute kernel is executed. Note that the results are not copied back to the host to prevent the pure CUDA implementation from blocking. We incrementally apply optimizations and show their effect.  As shown in Fig.~\ref{fig:dgemm_benchmarks}, the baseline bar indicates that the framework adds some overhead on top of the naive CUDA implementation, which is significant, especially for smaller matrices. PREMA overcomes some of these overheads with the optimizations presented. Note that many optimizations applied automatically by the heterogeneous framework could also have been implemented directly in the CUDA implementation. However, it would require much more effort from the user and a relatively more extensive and more complex application code. Moreover, the code would need to be rewritten to run on non-NVIDIA devices. 

\subsubsection{Page-locked Host Memory}
To fully utilize the bandwidth capabilities of the respective hardware, NVIDIA GPUs require that host data to be transferred to the GPU should reside in a page-locked memory region. Moreover, this is the only way that device-to-host transfers can be asynchronous with respect to the host. Thus, applications need to explicitly (de)allocate memory in a unique way, which increases code complexity and induces an overhead much higher compared to regular memory allocations. 
We incorporated this optimization into the framework to relieve applications and PREMA from explicitly handling this burden. For this purpose, the framework allocates a large chunk of page-locked memory at the initialization step that is later used as a memory pool for host memory allocations and prevents further expensive requests for page-locked host memory allocations. 

The optimization gives a significant boost that ranges between 30\% and 145\% for different matrix sizes (Fig.~\ref{fig:dgemm_benchmarks}; TF-PageLocked). Specifically, the smallest improvement is observed in smaller matrices (64x64) with 30\%, followed by the larger ones (768x768 and 1024x1024) with about 70\%. In the 128x128 case, it attains 90\%, while the most notable improvement with more than 100\% is achieved when 256x256 (120\%) and 512x512 matrices (145\%) are used. 

\subsubsection{Custom Device Memory Pools}
\label{sec::dev_mem_pool}
Allocating and freeing device memory are expensive operations that may require synchronization
between the host and the device. Moreover, the two functions might require the completion
of previously issued asynchronous operations before running. The proposed runtime system
uses a custom memory allocator per device to avoid overheads from constantly requesting new memory (de)allocations. During the initialization of the runtime system, a request to allocate most of the available memory of each device (except the host) is issued, and the custom memory allocator handles the returned memory. When memory needs to be allocated, the custom allocator is used instead of the one provided by the device library.

The most significant improvement of this optimization is manifested for the smallest matrix case (64x64) with another 60\%. As the size of the matrices increases, the improvement observed declines with 25\%, 12\%, and 6\%, respectively, for matrices ranging from 128x128 to 512x512. For larger matrices, the improvement is negligible (Fig.\ref{fig:dgemm_benchmarks}; TF-CustomAlloc).


\subsubsection{Enabling Concurrent GPU Operations} So far, all the optimizations implemented were focused on improving the overlap of the CPU and GPU operations; however, processes that run in the GPU are still serialized by default. We enable implicit concurrency  between the different GPU operations by utilizing multiple execution streams provided by the device implementations (OpenCL command queues or CUDA streams for NVIDIA GPUs). Our implementation uses multiple streams for submitting computation kernels and two for memory transfer operations (one for each direction). 
We found that five streams for computation kernels are generally enough to saturate the device capabilities for concurrent kernel executions. Still, we also provide an environmental variable that allows the application to change this without recompilation. 

The results of this optimization are substantial in most matrix sizes when the overlap between memory transfers and kernel executions can be large enough. This is the case for all matrix sizes larger than 64x64. Since the overhead of transferring data to the device is substantial, a great percentage of that can be overlapped with the kernel execution of the previous iteration. Thus, the results show improvements of 75\%, 50\%, 50\%, 85\%, and 100\%, respectively, for matrices of size 128x128-1024x1024. On the other hand, the improvement in the case of 64x64 is 10\% due to the small amount of data that needs to be transferred (Fig.\ref{fig:dgemm_benchmarks}; TF-TferQueue, TF-MultQueues). 
The final optimization applied in this aspect was to use more than one stream per data transfer direction. However, this did not improve performance further because the device hardware only supports two copy engines.

\begin{figure}
    \centering
    \includegraphics[width=.6\linewidth]{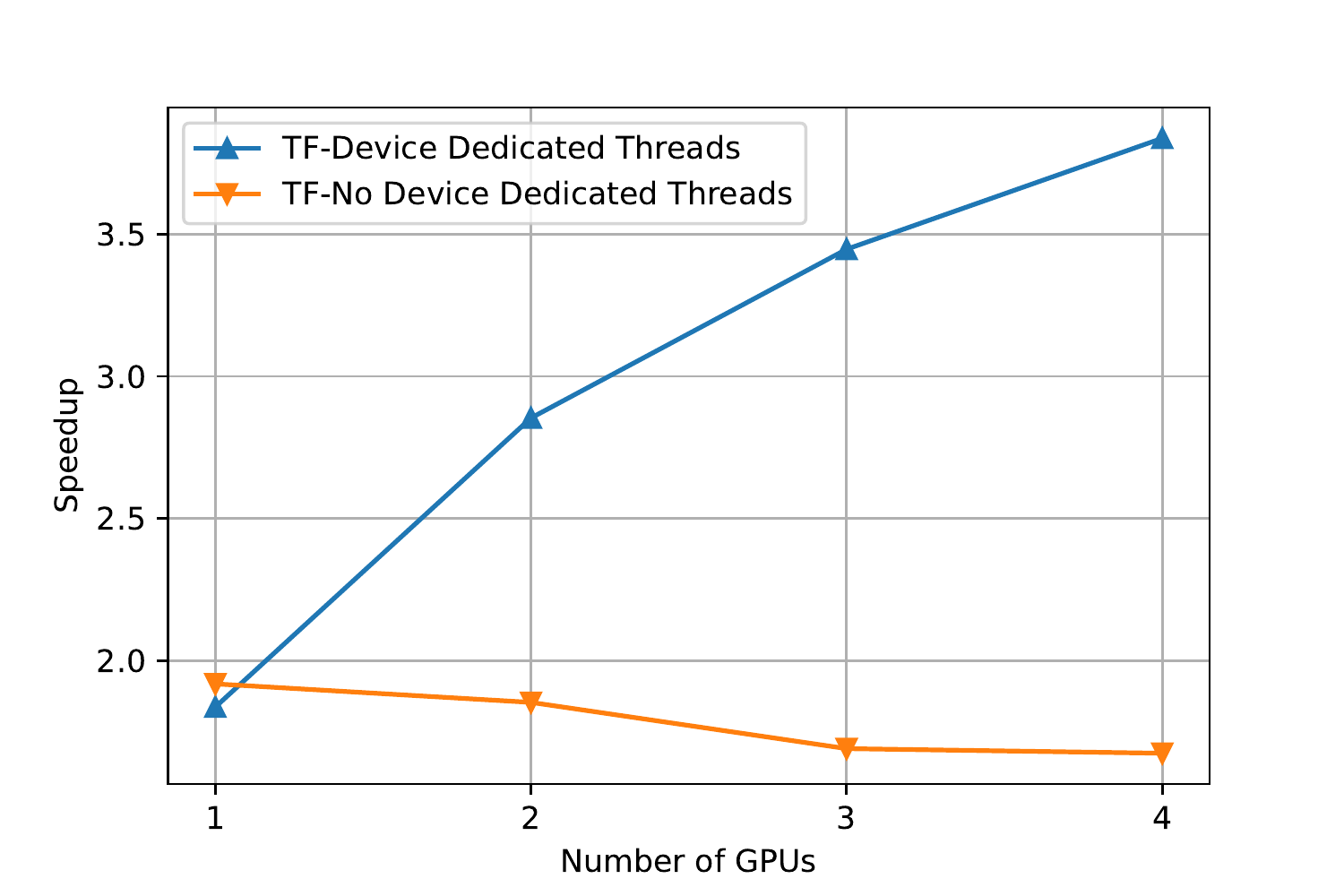}
    \caption{Performance of the heterogeneous tasking framework on a 64x64 matrix multiply benchmark utilizing multiple GPU devices with and without dedicated threads per device. The performance is shown as speedup against a simple CUDA implementation.}
    \label{fig:mult_dev}
\end{figure}

\subsubsection{Other optimizations} 
We have introduced request memory pools to mitigate the effects of system calls and thread synchronizations. Request pools are maintained per active thread, and the memory of a request is recycled in the pool once the respective operations have been completed. Another optimization regarding requests was implemented on the queues used to submit a request to a device. Initially, we used structures provided by the C++ STL, protected by a mutex, to implement such queues. These queues were substituted with custom lists that avoid allocating nodes to store new elements. Moreover, the mutex locking step was moved after the queue's size was checked to eliminate unneeded locking operations. This optimization is less important than the ones discussed above, about 2\% improvement, but helps to attain a more consistent latency, especially when the dedicated thread is used (Fig.\ref{fig:dgemm_benchmarks}; TF-Tpools).  

\subsubsection{Putting it all together}
To summarize, we have applied a series of optimizations that affected the performance of the tasking framework on the specific benchmark in different percentages depending on the size of the matrices. The most important optimizations include the automatic use of page-locked host memory and the introduction of multiple streams. However, the custom device memory pool also substantially improved the cases of smaller matrices. The overall performance improvement achieved from this series of optimizations can exceed 300\%, depending on the size of the matrices. Our framework offers all these optimizations with minimal user involvement and will continue to improve without any modifications required in the application code. Note that the performance improvements will increase as the number of memory transfers per computation decreases.

\subsubsection{Multi-Device Platforms}
The above results show the performance of the heterogeneous tasking framework on a single device. However, our framework is able to utilize multiple devices automatically. This is where the importance of using dedicated threads per device becomes apparent. While dedicated threads do not help when only a single GPU is in use (as shown in Fig.~\ref{fig:mult_dev}), they are crucial to scale beyond a single device. Fig.~\ref{fig:mult_dev} shows that the framework can scale almost linearly as we add more devices, achieving up to 3.8 speedup on 4 GPUs. The superlinear speedup observed in the case of one and two GPU devices is an effect of the optimizations presented so far for a single device. As we add more devices, the effect of these optimizations declines, and the speedup becomes linear.  We must note that the hardware available to us constraints the framework's capabilities. All the devices in the specific machine share a single bus with the host. Thus, the maximum memory transfer throughput is constant whether one or four devices are utilized. For the specific benchmark, a maximum of 2 64x64 matrices can be moved simultaneously to any device. Given a more capable hardware, we expect the framework to perform much better in a multi-device context.

\subsection{Heterogeneous PREMA}

\begin{figure}[t]
    \centering
\subfloat[Latency for small messages(left) and overall(right).]{
    \centering
    \includegraphics[width=0.49\linewidth]{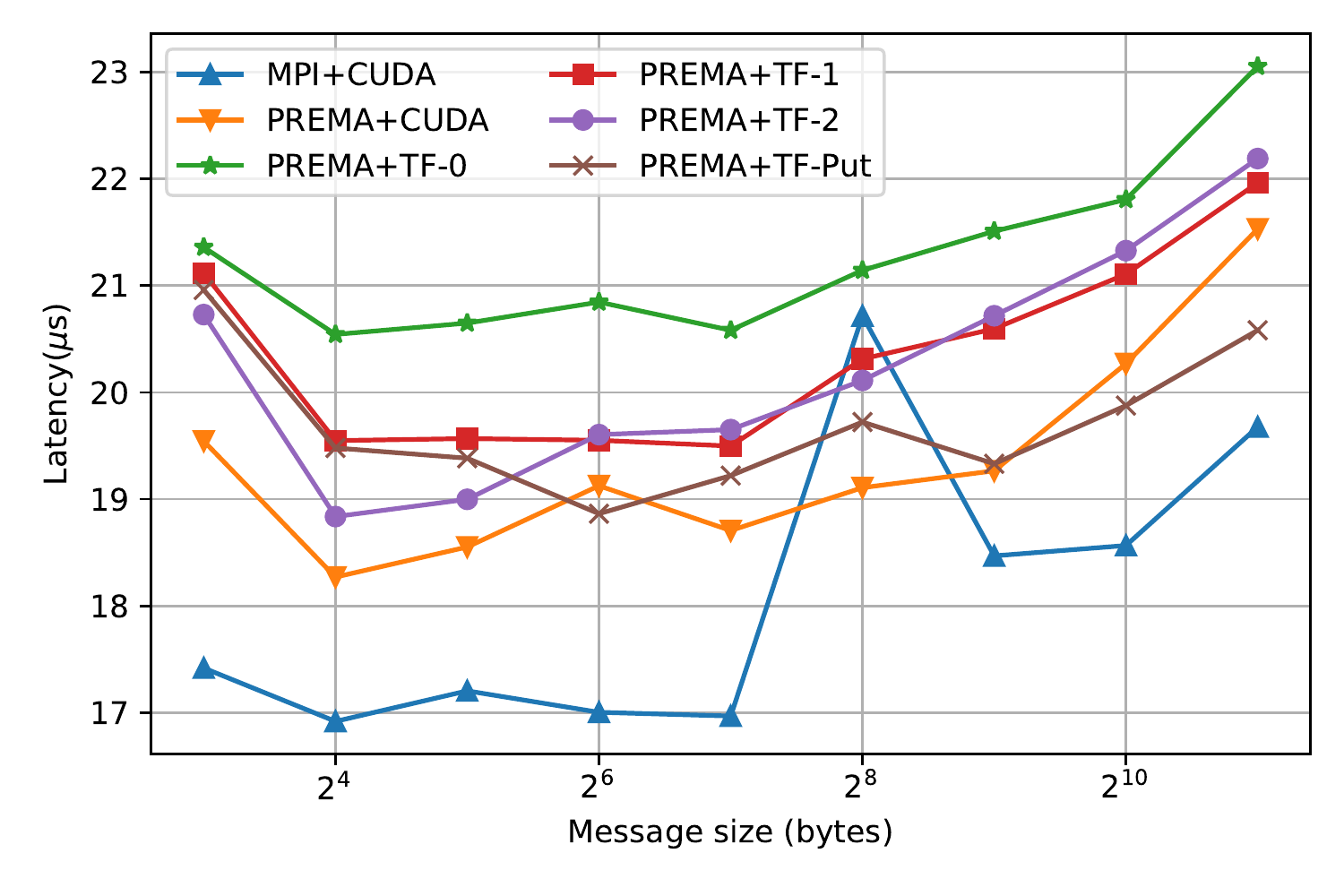}
    \includegraphics[width=0.49\linewidth]{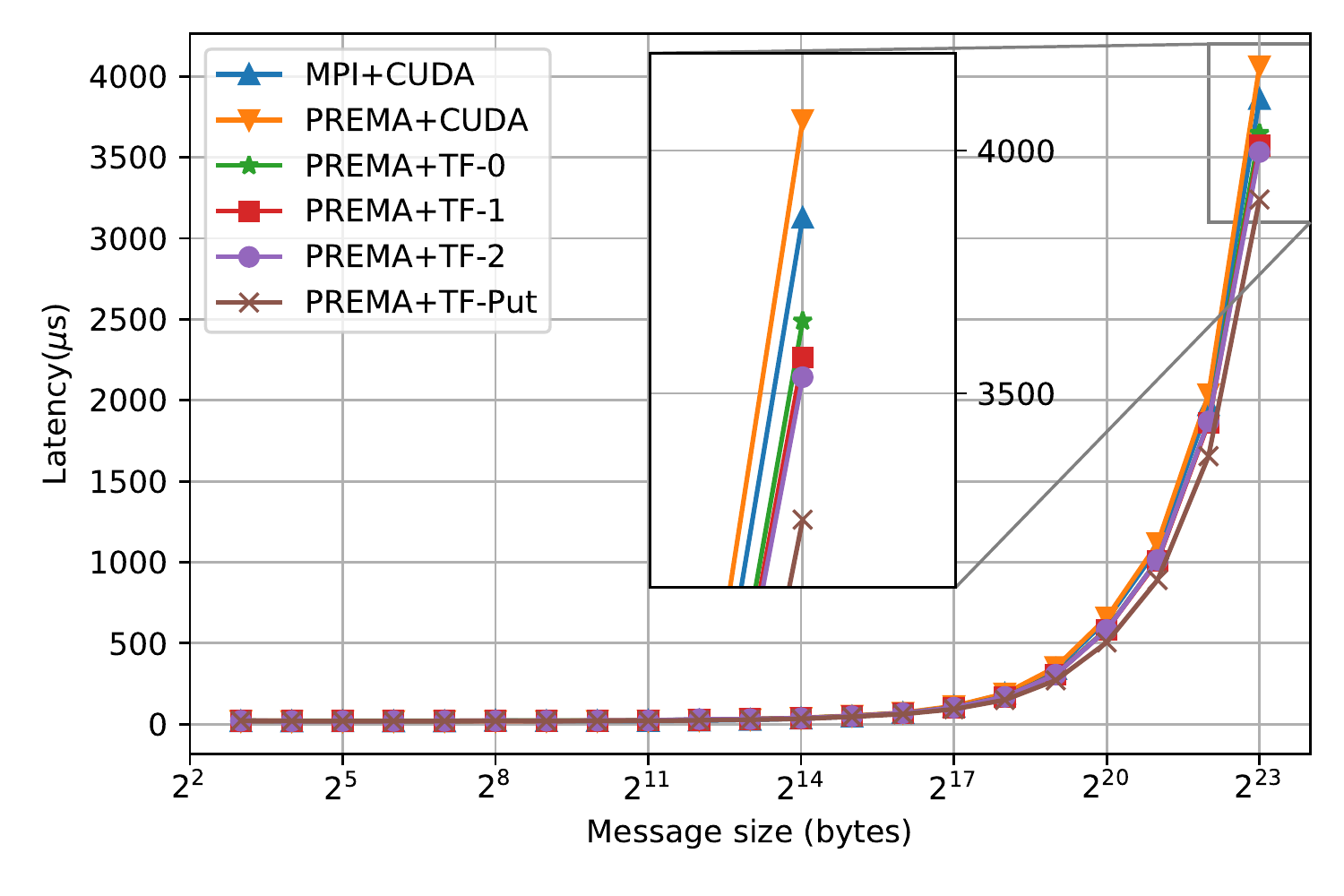}
    \label{fig:prema_opt_lat}
}
\\
\subfloat[Bandwidth for small messages(left) and overall(right).]{
    \includegraphics[width=0.49\linewidth]{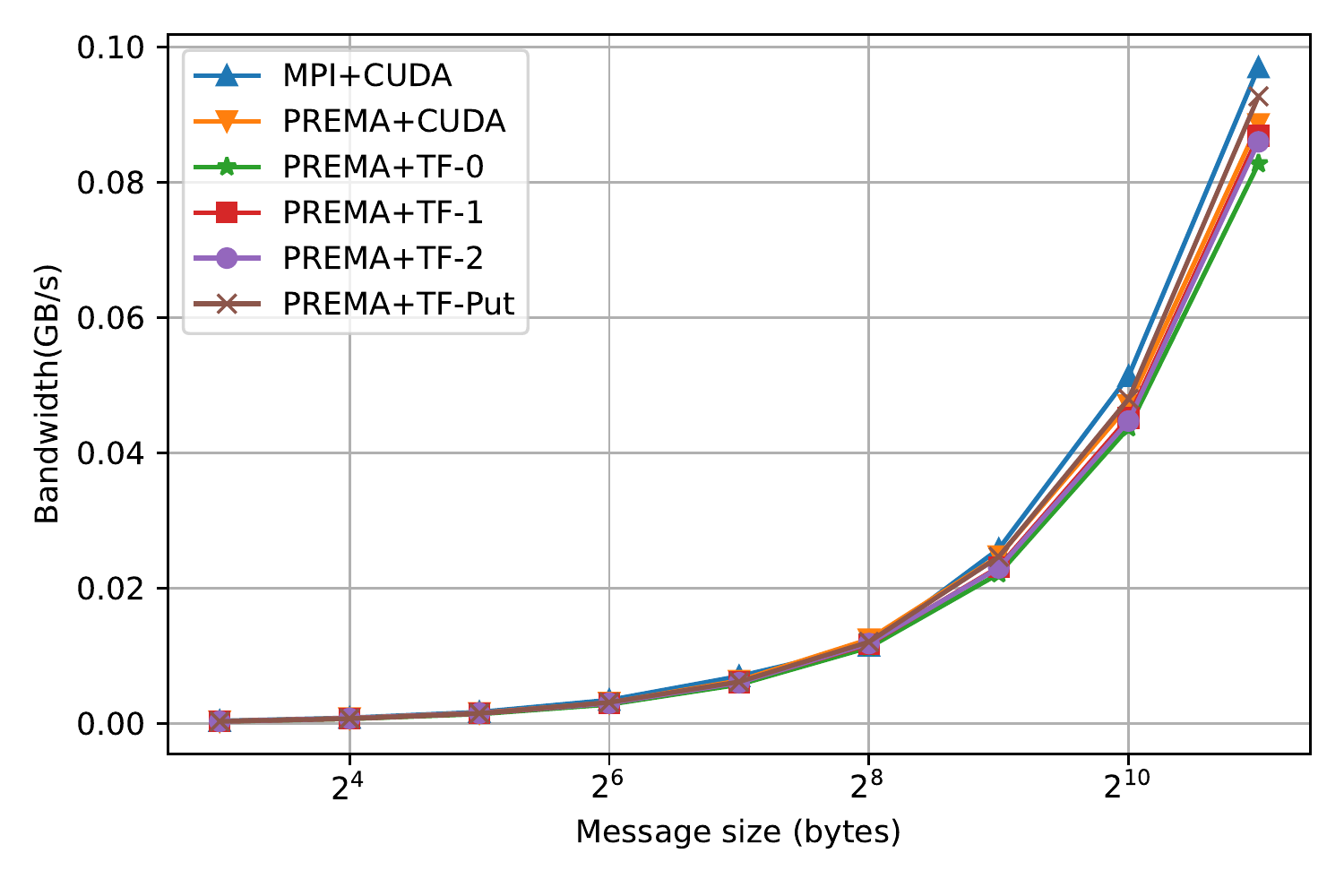}
    \includegraphics[width=0.49\linewidth]{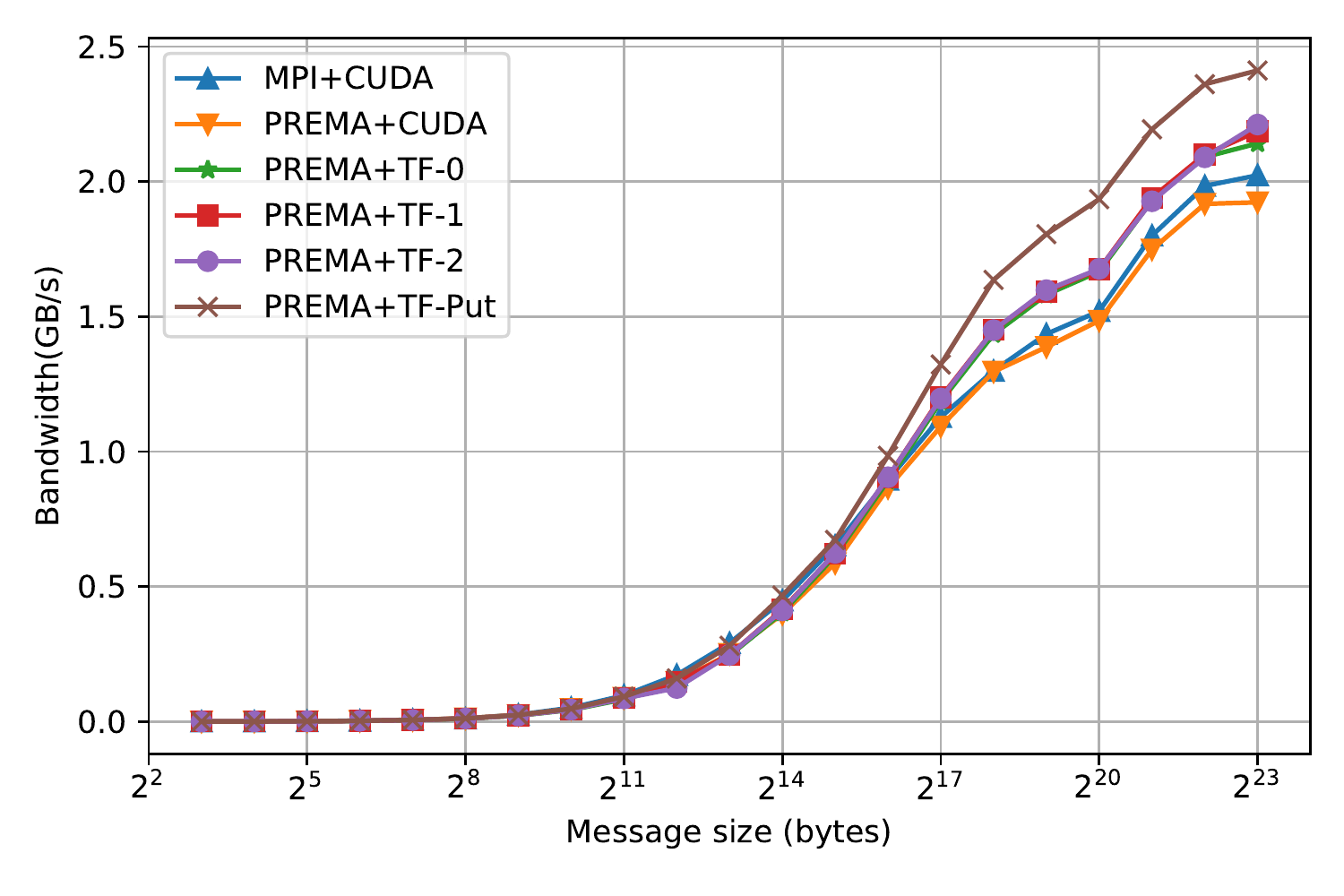}
    \label{fig:prema_opt_bw}
}
\caption{Performance evaluation of different optimization techniques in terms of (a) latency and (b) bandwidth on a ping-pong benchmark without direct network device-to-device transfers. Optimizations are compared against the baseline performance of an MPI+CUDA implementation. \textbf{PREMA+CUDA}: Integration of PREMA with CUDA directly, without involving the heterogeneous tasking framework.
\textbf{PREMA+TF-0}: The baseline integration of PREMA and the heterogeneous tasking framework. \textbf{PREMA+TF-1}: PREMA+TF-0 plus the introduction of memory pools. \textbf{PREMA+TF-2}: PREMA+TF-1 plus the optimization of avoiding message buffer copies for very small messages. \textbf{PREMA+TF-Put}: The remote put operation after the optimizations of PREMA+TF-2.}
\label{fig:prema_opt}
\end{figure}

To optimize the performance of the new heterogeneity-aware version of PREMA, we experiment with optimizations that can help us mitigate the overheads following the implementation of remote handler invocations that include heterogeneous memory both without and with the tasking framework (without a dedicated thread). We evaluate our optimizations on a simple ping-pong benchmark for inter-node communications and compare it with an MPI+CUDA implementation. The benchmark runs 100 ping-pong iterations with message sizes ranging between 8 bytes to 8 MBs, and the average latency and bandwidth observed per message size are reported.

\subsubsection{Device Message Receiving Cache}
In PREMA, messages are one-sided and asynchronous and are received implicitly to invoke a designated task on their target. Thus, the receiver cannot specify a memory region where the message buffer shall be stored (like MPI). PREMA has to dynamically allocate memory to receive the incoming message buffer and provide it to the application handler invocation. In our initial implementation, without the tasking framework, simply allocating new device memory for each incoming message resulted in poor performance, increasing the latency experienced up to ten times compared to the respective MPI implementation. We avoided this behavior by allocating a cache in the device specifically for the buffers of received messages. When a new message buffer is about to be received, memory is requested from the cache instead of the device API if possible. The cache allowed us to attain performance within 10\% overhead of that achieved by the MPI (Fig.~\ref{fig:prema_opt}; PREMA+CUDA). It is also important to note here the consistent ``spike'' observed in the MPI performance for messages of size 256B, which does not seem to affect our implementation when using the device cache. 

\subsubsection{Preallocating hetero\_objects}
Hetero\_objects automatically utilize memory pools for device memory, thus, implicitly overcoming the issue faced in the case where device memory is handled explicitly and achieving performance within 25\% of the MPI+CUDA implementation (Fig.~\ref{fig:prema_opt}; PREMA+TF-0). However, we can still improve some latencies caused by the constant allocation and deallocation of temporary hetero\_objects that wrap message buffers targeting device memory. Specifically, we found that the data structures allocated for a hetero\_object for bookkeeping different operations targeting the object in various devices can significantly affect communication performance. Since these structures can be allocated in advance, we use a pool of preallocated semi-initialized hetero\_objects to mitigate this effect. 
This optimization improved the latency experienced for small messages by about 10\%, bringing the performance of PREMA within 15\% overhead of MPI+CUDA, as can be seen in Fig.~\ref{fig:prema_opt} (PREMA-TF-1). Moreover, for larger messages the performance achieved is better from the MPI+CUDA by almost 10\% as can be seen when zooming in the latency of 8MB messages. 

\subsubsection{Avoiding Message Buffer Copies}
The process of inter-node message transfers presented in \ref{sec:hetero_prema_impl} with host-staging includes an optimization for small buffer sizes to only send one message. Small messages with a size up to 512 bytes (header + buffer size) will consist of the actual application data/buffer appended at the end of the message. This helps optimize the performance of small messages where even negligible overheads are noticeable. However, when hetero\_objects are used, this introduces an extra copy. As explained before, requesting access to the underlying data of a hetero\_object will implicitly copy the data to the host in a buffer maintained by the framework. To append the data at the end of the message header, PREMA needed to copy the data from the framework's location to the message header.

A new method is introduced in the hetero\_object API that allows the user to request a copy of its underlying data to a designated memory region at the host. PREMA utilizes this feature to request from the framework to directly transfer the device data at the end of the message header buffer. This change slightly improves the communication critical path and attains up to 5\% lower latency for smaller messages (Fig.~\ref{fig:prema_opt}; PREMA-TF-2). 

\begin{figure}
    \centering
\subfloat[Latency for small messages(left) and overall(right)]{

    \includegraphics[width=0.49\linewidth]{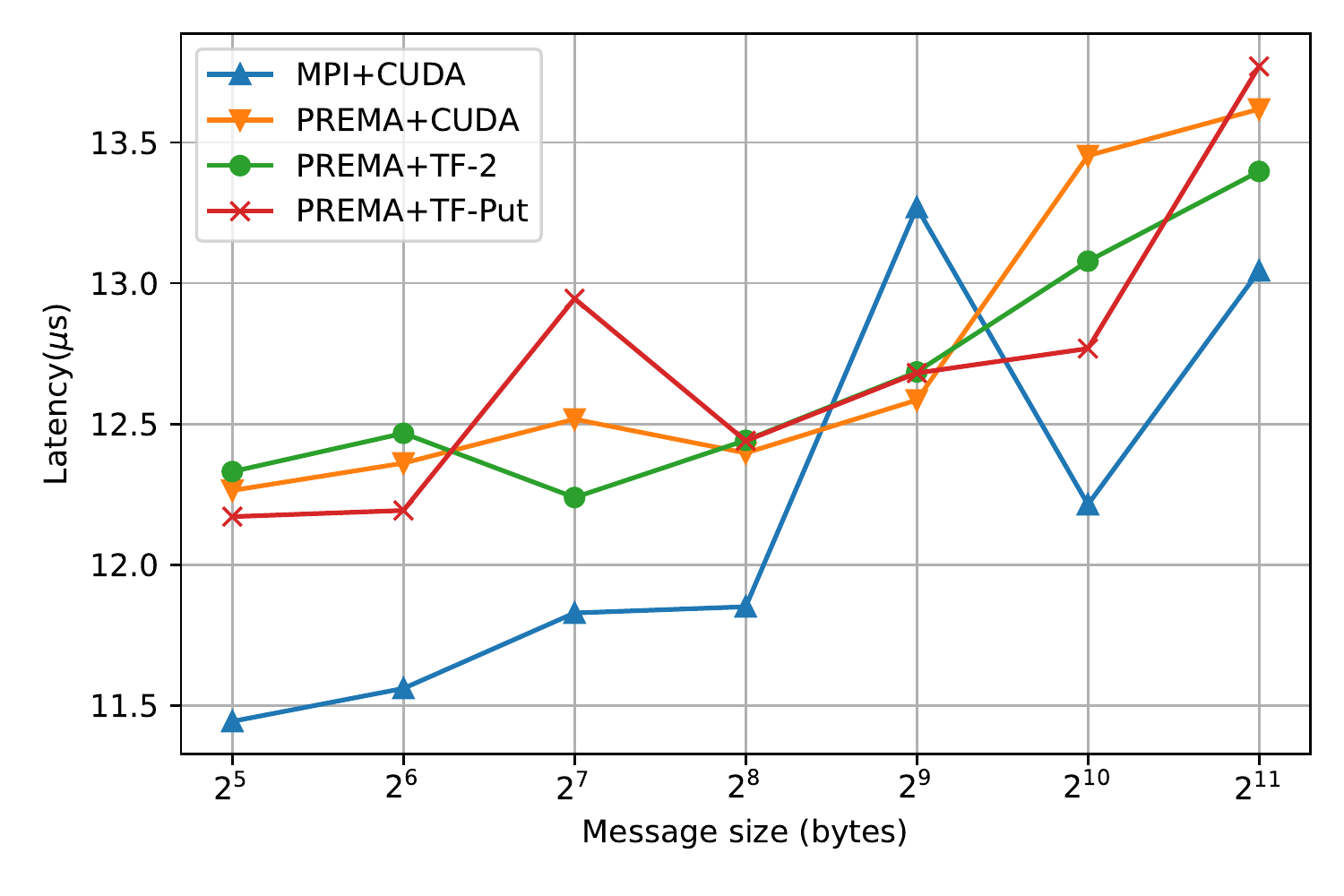}
    \includegraphics[width=0.49\linewidth]{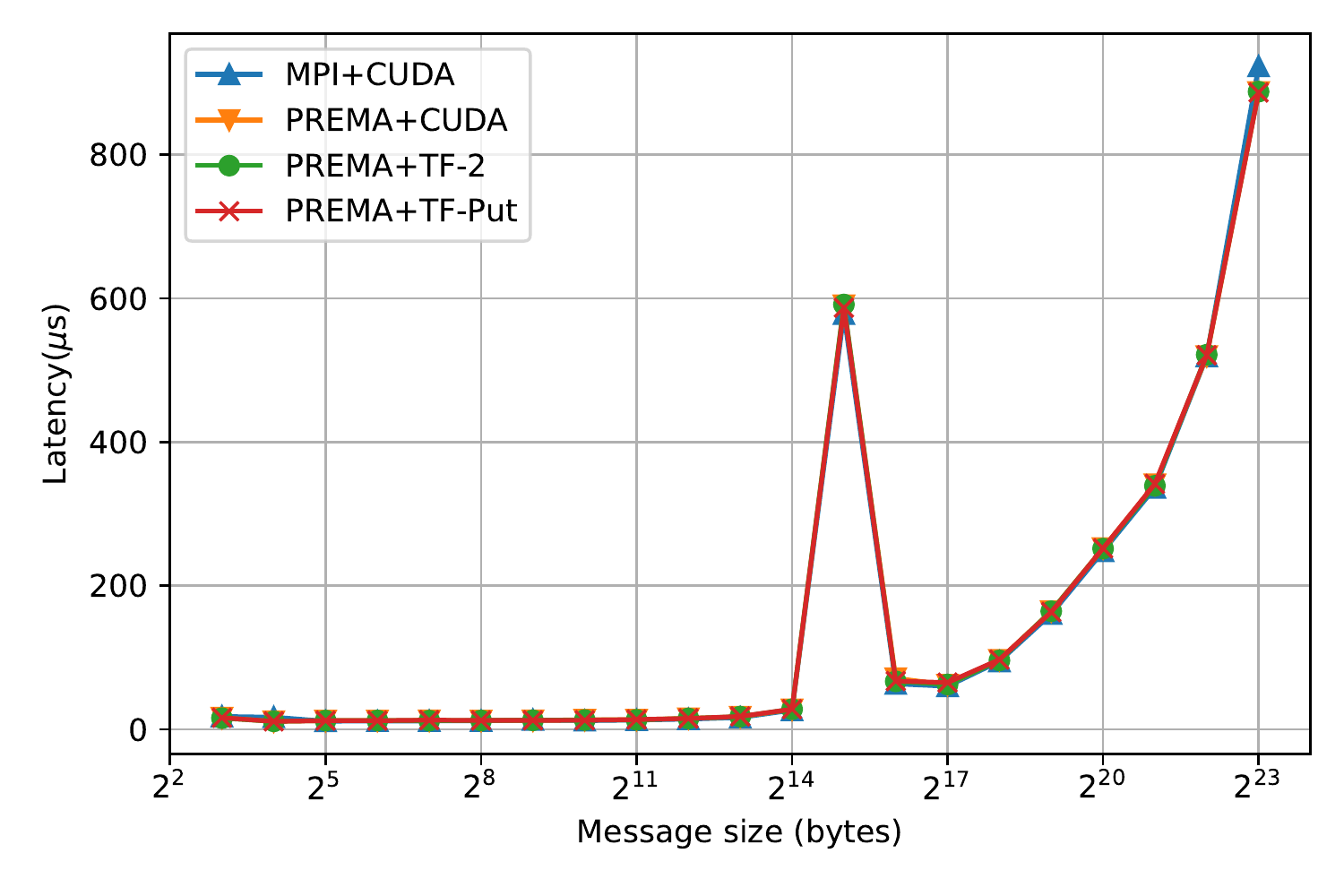}
    \label{fig:prema_opt_lat_dd}
}
\\
\subfloat[Bandwidth for small messages(left) and overall(right)]{

    \includegraphics[width=0.49\linewidth]{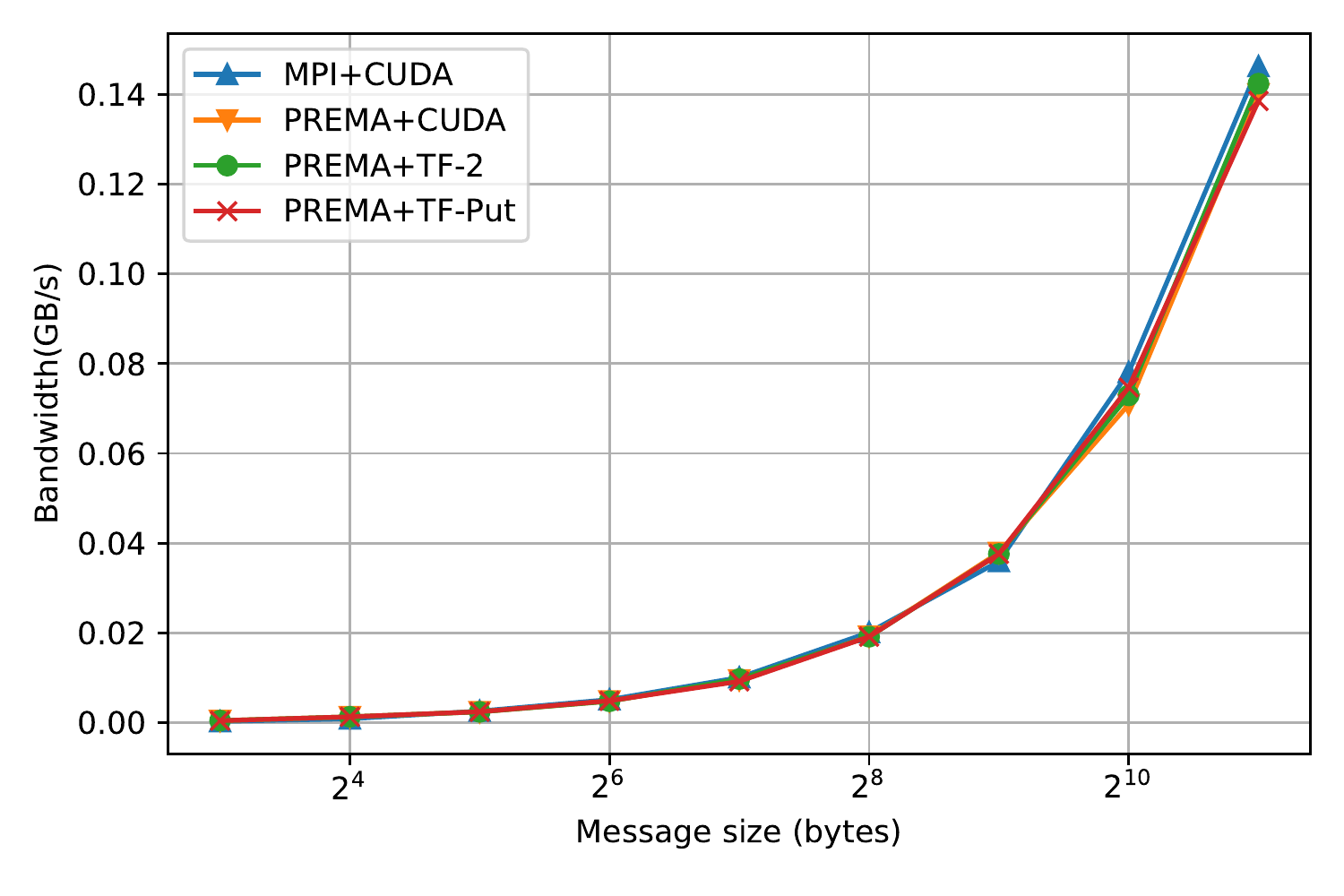}
    \includegraphics[width=0.49\linewidth]{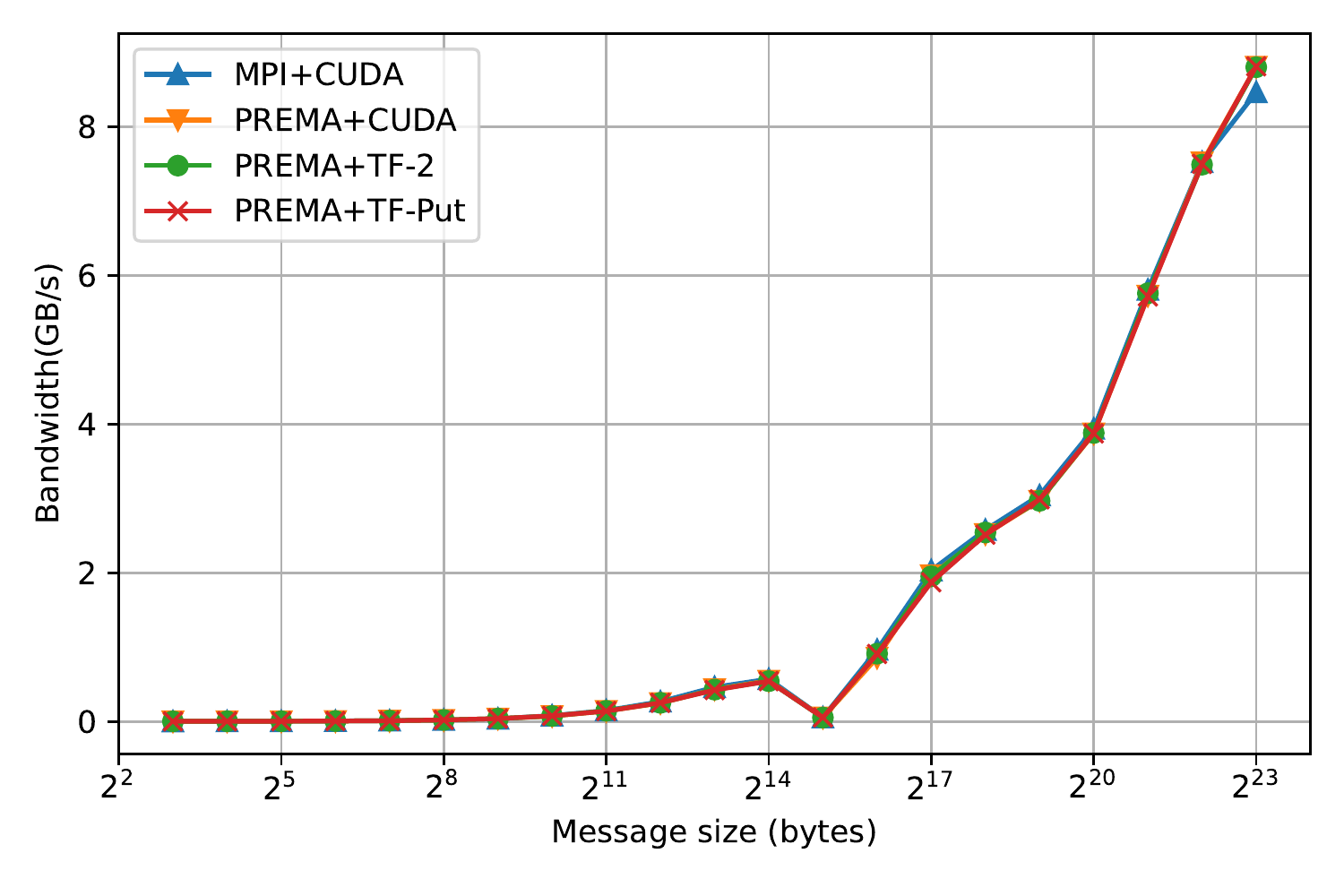}
    \label{fig:prema_opt_bw_dd}
}
\caption{Performance evaluation of different optimization techniques in terms of (a) latency and (b) bandwidth on a ping-pong benchmark with direct network device-to-device transfers. Optimizations are compared against the baseline performance of an MPI+CUDA implementation.  \textbf{PREMA+CUDA}: Integration of PREMA with CUDA directly, without involving the heterogeneous tasking framework.
\textbf{PREMA+TF-2}: Integration of PREMA and the heterogeneous tasking framework incorporating all the optimizations presented in this section. \textbf{PREMA+TF-Put}: The remote put operation after the optimizations of PREMA+TF-2.}
\label{fig:prema_opt_dd}
\end{figure}

\begin{figure}
    \centering
\subfloat[Latency for small messages(left) and overall(right)]{

    \includegraphics[width=0.49\linewidth]{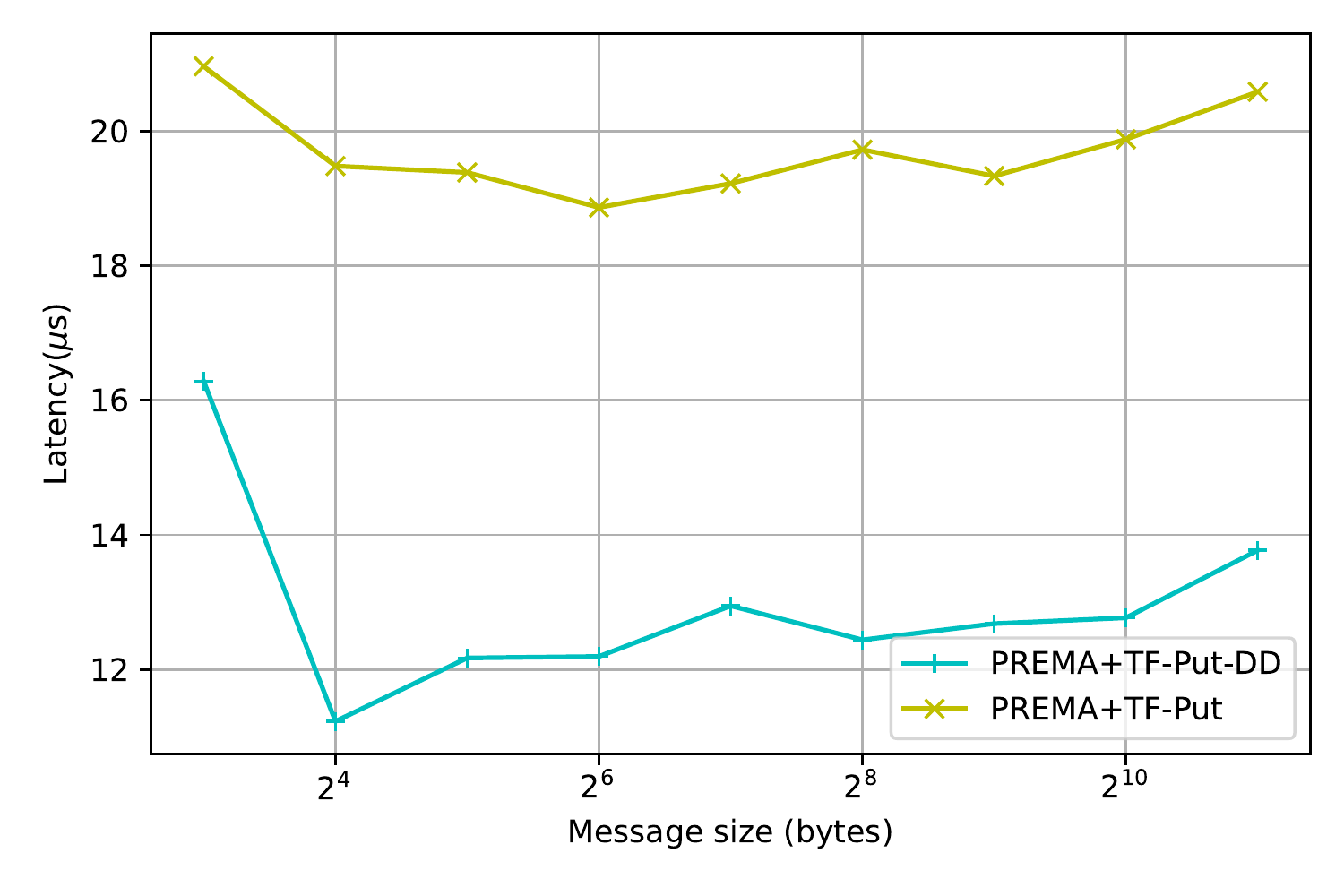}
    \includegraphics[width=0.49\linewidth]{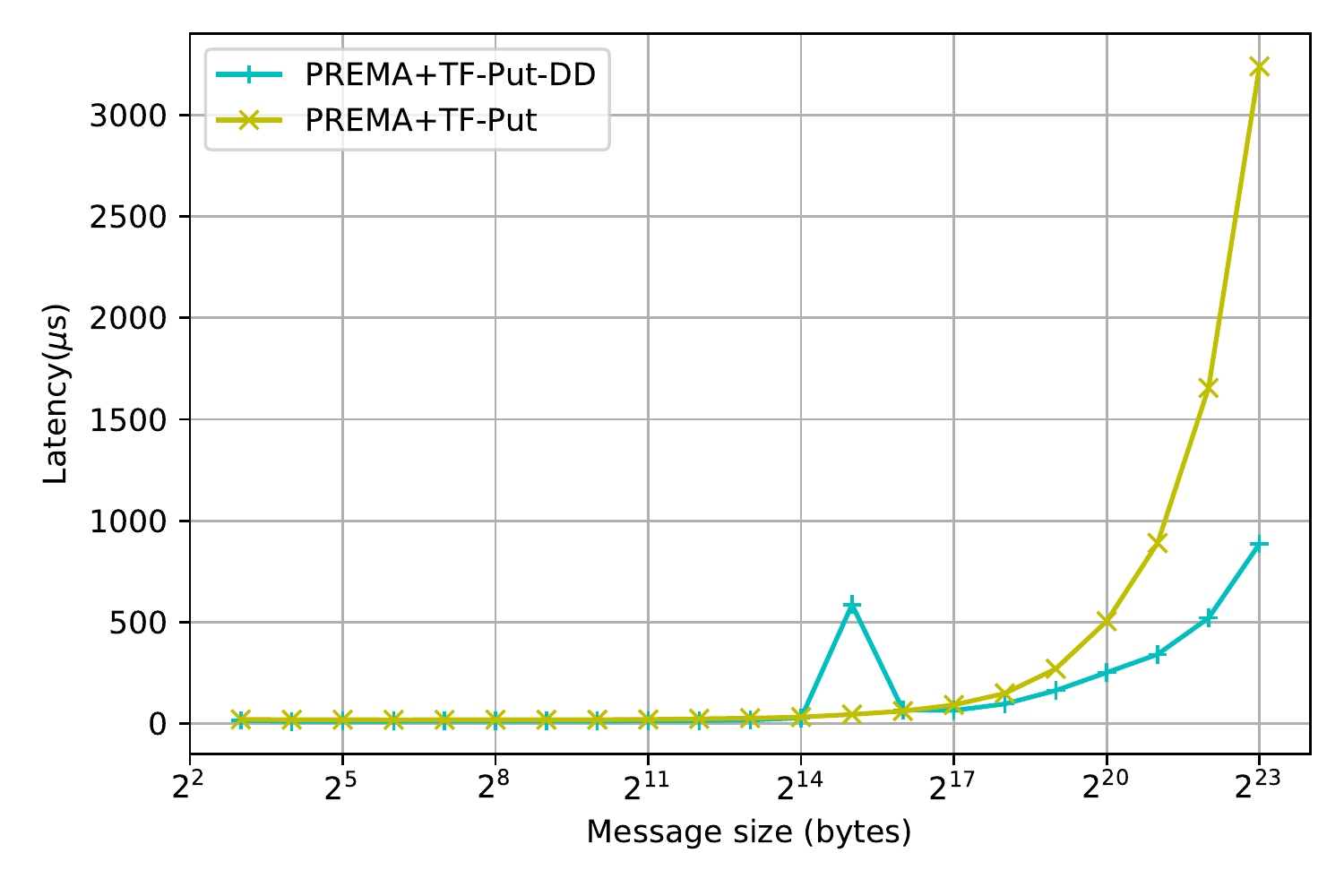}
    \label{fig:stage_vs_dd_lat}
}
\\
\subfloat[Bandwidth for small messages(left) and overall(right)]{

    \includegraphics[width=0.49\linewidth]{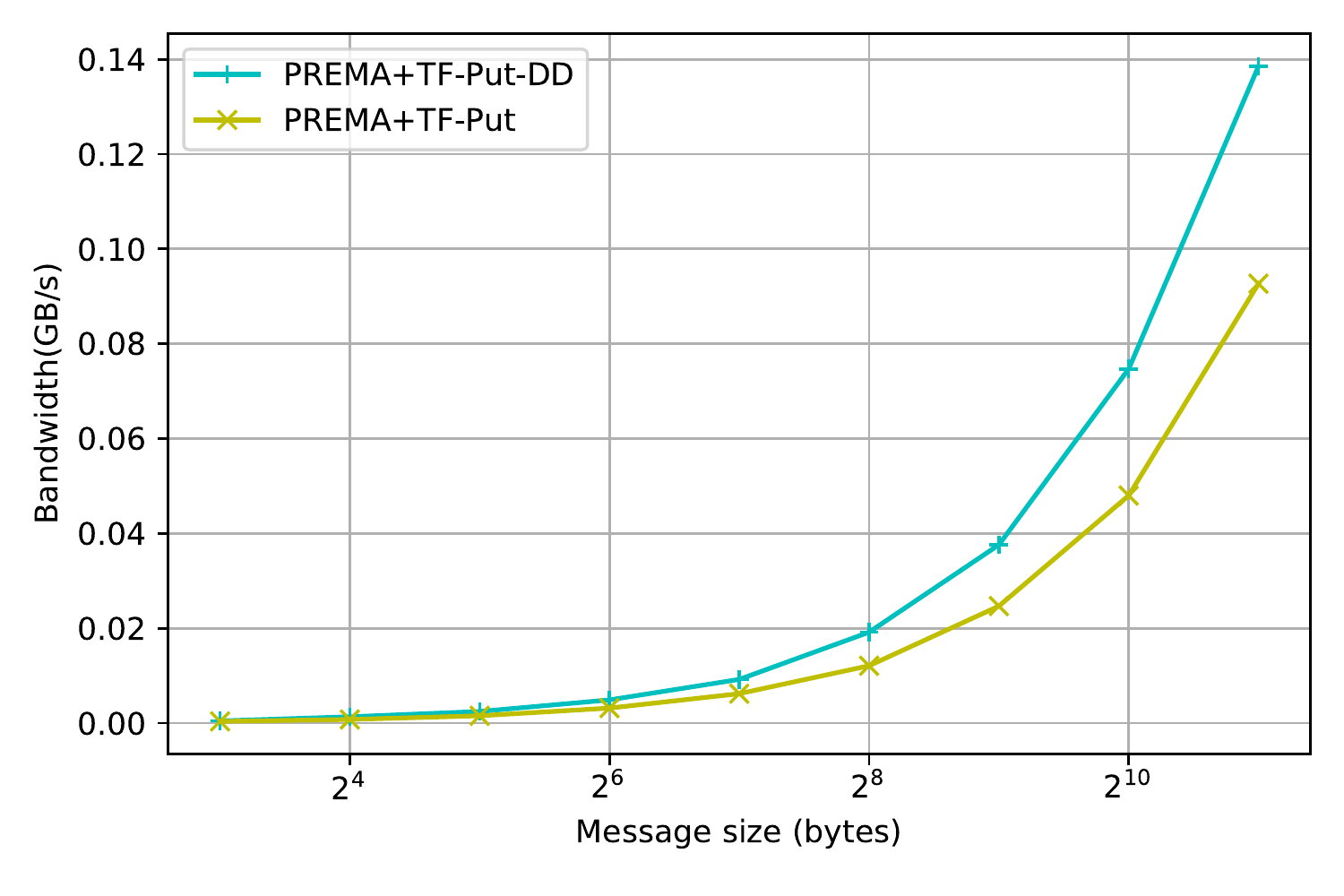}
    \includegraphics[width=0.49\linewidth]{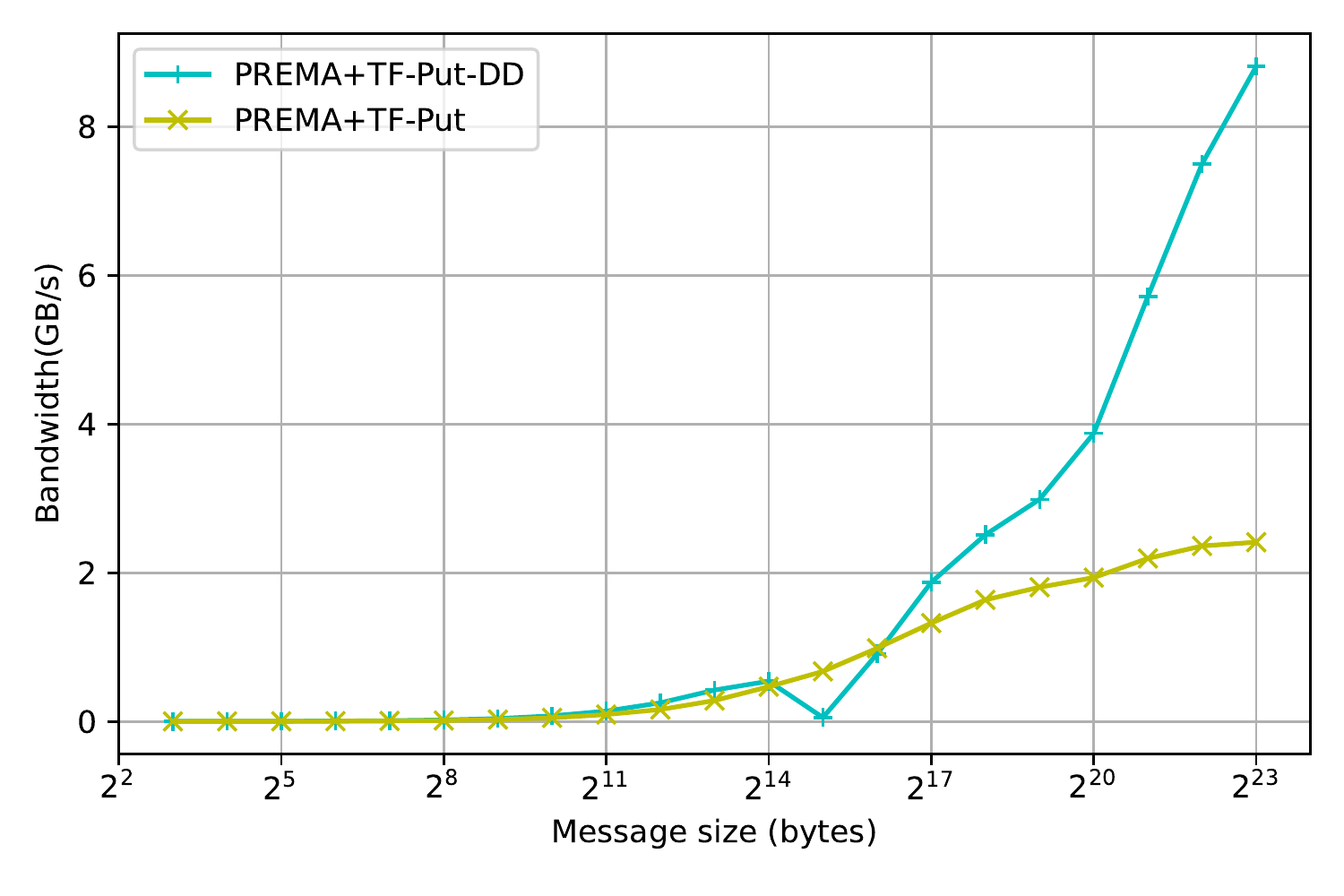}
    \label{fig:stage_vs_dd_bw}
}
\caption{Comparison of the optimal performance in terms of (a) latency and (b) bandwidth using host-staging or Device-to-Device (DD) network transfers for the ping-pong benchmark.}
\label{fig:stage_vs_dd_comparison}
\end{figure}

\subsubsection{Put Operation}
Another operation introduced to further increase the performance of inter-device communication over the network is the put operation. As mentioned earlier, the put operation allows PREMA to utilize the existing memory of heterogeneous objects that is also page-locked. By leveraging these optimizations, the put operation outperforms all previous optimizations that use the tasking framework since the transfer from the host to the device is much faster on the receiver side. Fig.~\ref{fig:prema_opt}(PREMA-TF-Put) shows its performance reaching and overcoming the one of the PREMA+CUDA implementation 
 for messages larger than 64 bytes and even the MPI+CUDA for messages larger than 8KB achieving up to 20\% better performance for messages of 8MB. 

\subsubsection{Direct to Device Transfers}
The optimizations presented so far target the generic implementation of heterogeneity on top of PREMA, where the communication library/hardware is not heterogeneity-aware. However, as mentioned in detail in \ref{sec:hetero_prema_impl}, PREMA can leverage the capabilities of hardware/libraries that have been integrated with support for direct device-to-device communication. Following the previously presented procedure, the latencies observed for heterogeneity-aware hardware can be significantly mitigated. Fig.~\ref{fig:prema_opt_dd} shows the performance of the optimized version of each operation and the MPI+CUDA when direct-to-device communication is possible. The attained performance, in this case, is up to 100\% better than the host-staging case for small messages and up to 200\% for large messages, as shown in Fig.~\ref{fig:stage_vs_dd_comparison}. 

\subsubsection{Summary}
Overall, all three operations introduced in PREMA to handle heterogeneity, including the transfer of CUDA  buffers, hetero\_objects, and the remote put operation, significantly simplify application development, saving hundreds of lines of boilerplate code while maintaining reasonable overhead (10-15\%) compared to MPI+CUDA regarding latency and bandwidth. Moreover, it is interesting to notice that some of these operations, like the put operation, outperform MPI+CUDA (by up to 20\%) for large messages (Fig.~\ref{fig:prema_opt},~\ref{fig:prema_opt_dd})  by implicitly leveraging from the page-locked memory of hetero\_objects. Page-locked memory forces the operating system to lock a virtual memory address to a specific physical address, allowing CUDA GPUs to use DMA and significantly improve the throughput of read and write operations.

\subsection{Proxy Application: Jacobi3D}
\label{sec:jacobi3d}
\begin{figure}
    \centering
    \subfloat[Jacobi3D Strong Scaling]{
        \includegraphics[width=0.495\linewidth]{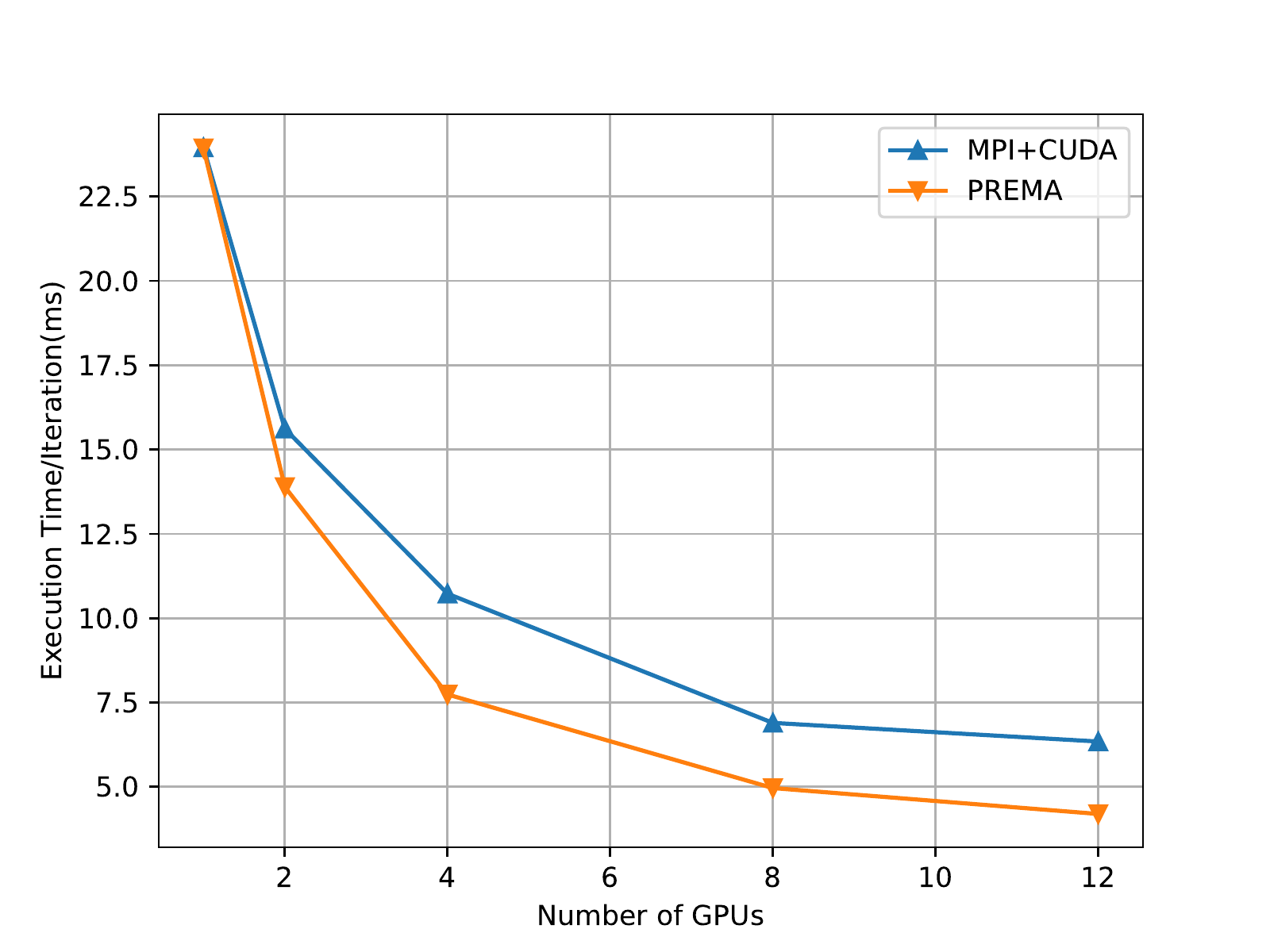}
        \label{fig:jacobi3d_strong}
    }
    \subfloat[Jacobi3D Weak Scaling]{    \includegraphics[width=0.495\linewidth]{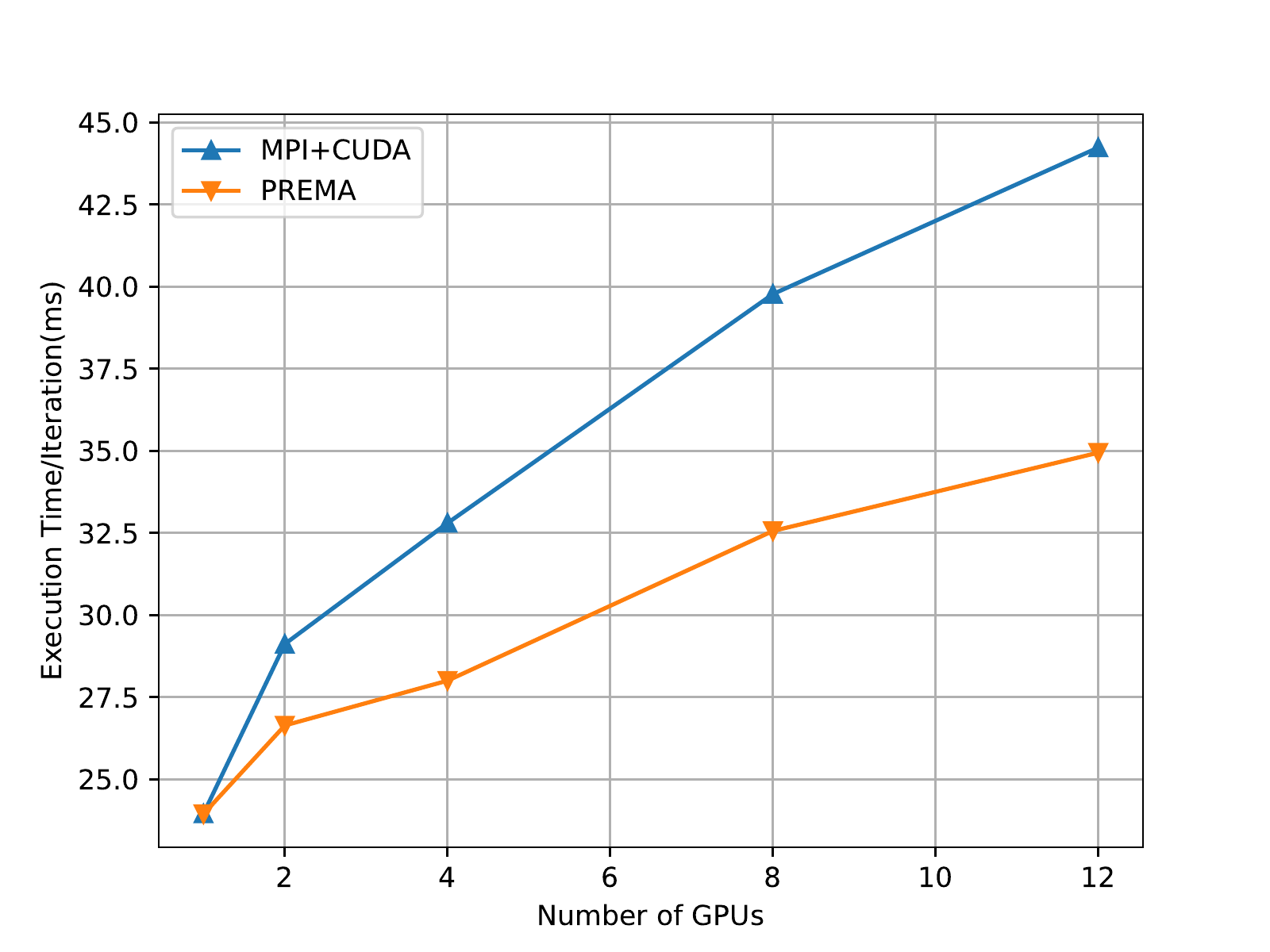}
    \label{fig:jacobi3d_weak}
    }
    \caption{Strong (left) and weak (right) scaling performance of the Jacboid3D proxy application.}
    \label{fig:ev_jacobi3d}
\end{figure}
To evaluate the performance of the distributed memory framework, we have adapted a proxy Jacobi3D application~\cite{jacobi3dCharm} on PREMA. The proxy performs a fixed number of iterations of the Jacobi method on GPUs in a 3D domain decomposed into cuboids and wrapped into mobile objects. In each iteration, the mobile objects exchange halo data, packing the GPU data and transferring them to their neighbors. On the receiving side, the data are unpacked into the GPU, and once all halos have been received, the Jacobi update is executed. 

Fig.\ref{fig:ev_jacobi3d} (left) shows the execution time of the heterogeneous PREMA versus the MPI+CUDA counterpart for a 1024x1024x768 domain (strong scaling). The implementation with PREMA achieves up to 23\% better performance. For weak scaling, the domain's size is increased according to the number of distributed GPUs.  The performance achieved is up to 25\% (Fig.\ref{fig:ev_jacobi3d}; right) better than the MPI+CUDA implementation. The improvements observed stem from automatically overlapping message passing, host-device memory transfers, and kernel invocations.

\subsection{Over-decomposition}
\begin{figure}
    \centering
    \includegraphics[width=\linewidth]{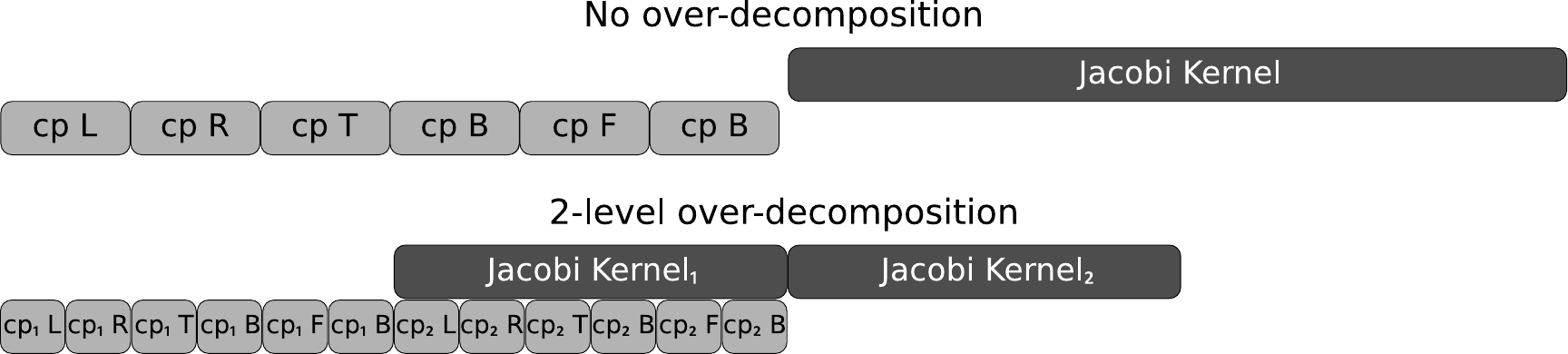}
    \caption{An example of the Jacobi3D execution timeline with no and 2-level over-decomposition. Each kernel invocation requires the copy of the six halo buffers, left (L), right (R), top (T), bottom (B), forward (F), and back (B), into the GPU. By decomposing the data domain (and the kernel invocation) into two pieces, PREMA can overlap the memory transfers of the second with the kernel invocation of the first, saving a significant amount of time.}
    \label{fig:timeline_comparison}
\end{figure}
\begin{figure}
    \centering
    \subfloat[Strong Scaling with Over-Decomposition]
    {
        \includegraphics[width=.495\linewidth]{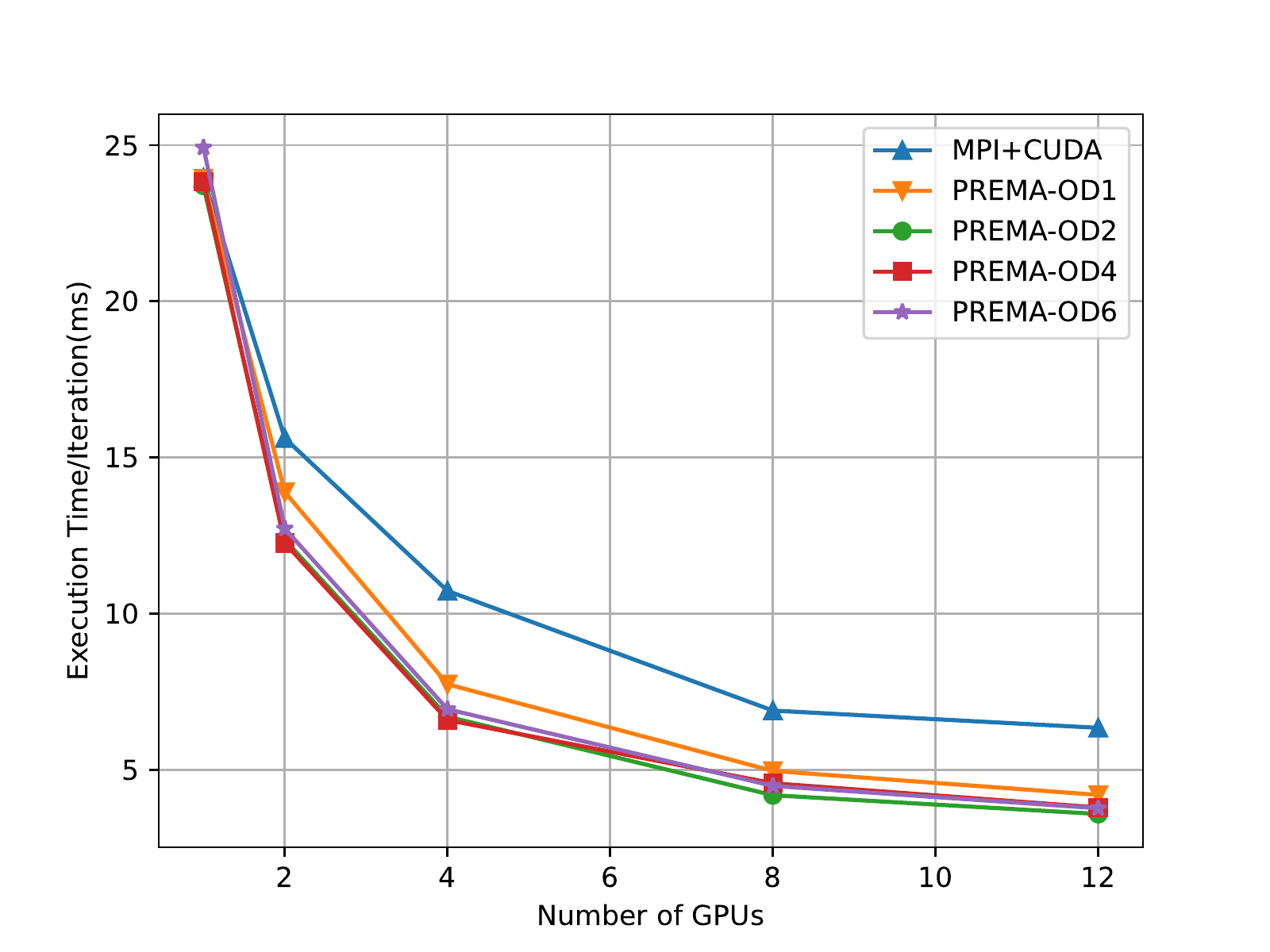}
        \label{fig:jacobi3d_od_strong}
    }
    \subfloat[Weak Scaling with Over-Decomposition]
    {    
        \includegraphics[width=.495\linewidth]{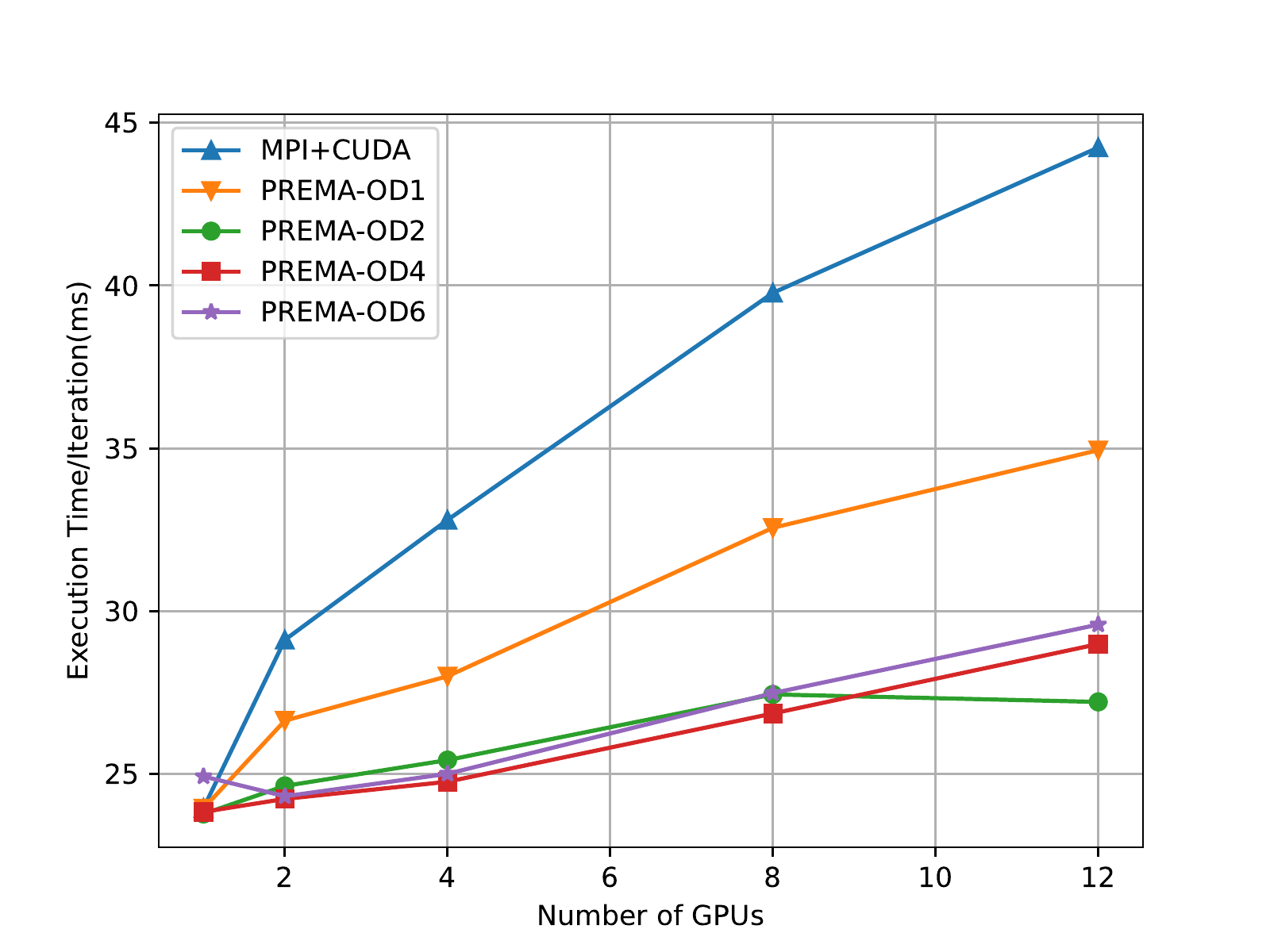}
        \label{fig:jacobi3d_od_weak}
    }
    \caption{Strong (left) and weak (right) scaling performance of the Jacboid3D proxy application with different levels of over-decomposition.}
    \label{fig:ev_jacobi3d_od}
\end{figure}

A common practice that PREMA applications utilize for performance improvement is over-decomposition. Over-decomposition is used to decompose the data domain into more chunks than the number of PEs, allowing PREMA more flexibility to load balance workload and overlap latencies. The effectiveness of this approach has already been demonstrated in previous work for heterogeneous platforms~\cite{Thomadakis22M, Thomadakis23T}. In the context of heterogeneity, host-to-device, and device-to-host memory transfers are broken into pipelined pieces and can be overlapped much more easily with the following kernel invocations. An example of the execution timeline that is achieved can be seen in Fig.~\ref{fig:timeline_comparison}. The effects of over-decomposition are shown in Fig.~\ref{fig:ev_jacobi3d_od} for the same Jacobi3D proxy application. Different levels of over-decomposition are attempted in this benchmark, with each level attaining improvements over the MPI implementation as well as the PREMA implementation without over-decomposition (OD1). The best performance is observed with an over-decomposition of two, which achieves improvements of up to 40\% versus the MPI implementation and about 20\% over the initial PREMA implementation. 

\subsection{Discussion}
On top of the end-user productivity and ease of use provided by the abstractions introduced, the framework provides numerous quantitative enhancements. Overall, this work has presented the following improvements:
\begin{enumerate}
    \item Implicitly managing memory transfers and task execution dependencies across multiple hardware devices eliminates the need for code refactoring and decreases complexity.
    \item Computing performance is boosted by up to 300\% implicitly leveraging from hardware-specific optimizations on a single GPU.
    \item A configurable scheduler that optimizes data locality and evenly distributes workload across multiple devices allows seamless utilization of multi-GPU nodes and attains linear scalability. 
    \item The potential heterogeneity awareness of the underlying network interface is automatically utilized, eliminating the need for explicitly written code to take advantage of this feature, providing up to three times better performance.
    \item The portable abstractions for communication among distributed devices remove the need for explicit monitoring of device-host and inter-node transfers to ensure data consistency and completion. While this approach does incur some overheads within 10\% of MPI for small messages, it also delivers exceptional performance for large messages, with gains of up to 20\%.
    \item Leveraging from these optimizations, a proxy 3-dimensional Jacobi application on top of PREMA achieved performance improvements of up to 30\%.
    \item Introducing over-decomposition further increases the distributed Jacobi's performance by 40\% compared to the MPI+CUDA implementation by allowing implicit operation pipe-lining and latency overlapping.
\end{enumerate}

\section{Conclusion and Future Work}
\label{sec:conc}

This paper introduced an efficient tasking framework that handles performance portability for multi-device heterogeneous nodes. This framework automatically scales applications to multiple devices while managing efficient scheduling, load balancing, task dependencies, memory transfers, and overlapping latencies, attaining a performance improvement of up to 300\%. Integrating this framework into a distributed runtime system resulted in a complete library that can leverage exascale HPC systems consisting of multiple heterogeneous nodes. In addition, this paper presents a series of optimizations and their evaluation which indicates mitigated overheads to within 10\% of the MPI. Evaluation results on a proxy application show that the end product of this work incurs low latency and scalable performance (up to 40\% versus the MPI+CUDA) while providing a simple and uniform interface independent of the target hardware. In the future, we plan to improve the tasking framework's performance further and provide compiler support to generate device kernels automatically. Moreover, we intend to extend PREMA's implicit load balancing layer to accommodate the workload of heterogeneous devices and experiment with a range of load balancing and scheduling policies.

\section*{Declarations}
\subsection*{Ethical Approval}
Not applicable.

\subsection*{Competing Interests}
The authors declare no competing interests.

\subsection*{Authors' Contributions }
P.T. wrote to the main manuscript text. Both authors contributed to the conception and design of the work. P.T. developed the software and benchmarks of this work. Both authors reviewed the manuscript.

\subsection*{Funding}
This work is partly funded by the Dominion Fellowship, the Richard T. Cheng Endowment at Old Dominion University, and NSF grants: CCF-1439079, CNS-1828593.

\subsection*{Availability of Data and Materials }
The software developed as part of this work is not publicly available but will be in the future.

\bibliographystyle{ieeetr}
\bibliography{references}

\end{document}